\documentclass[11pt]{elsarticle}
\usepackage[a4paper,margin=1in]{geometry}

\usepackage[T1]{fontenc} 
\usepackage[normalem]{ulem}

\usepackage{moreverb}
\usepackage[nounderscore]{syntax}

\usepackage[cmex10]{amsmath}
\usepackage{amssymb}
\usepackage{mathtools}
\usepackage{amsthm}
\usepackage{amsfonts}

\usepackage{gensymb}

\usepackage{subfig} 
\usepackage[export]{adjustbox}

\usepackage{multirow} 
\usepackage{rotating} 
\usepackage{booktabs} 
\usepackage{colortbl} 
\usepackage{tablefootnote} 
\usepackage[pdftex,colorlinks=true,urlcolor=blue,citecolor=blue]{hyperref}
\usepackage{xspace}
\usepackage{pifont} 

\usepackage{csquotes}

\usepackage{enumitem}

\usepackage{balance}
\usepackage{tikz}
\usepackage{microtype}

\begin{document}

\begin{frontmatter}

\title{Quantum Integrated High-Performance Computing: Envisioning a Layered \\Architecture for Next-Generation Hybrid Computing Infrastructure}

\author[inst1,inst4,inst2]{Suman Raj\fnref{*}}
\fntext[*]{Suman Raj was a visiting student at d with primary affiliation at b, when the work in this paper started. Her current affiliation is a. Siva Sai was also a visiting student at d, and his current affiliation is c. \\Correspondence to: School of Computing and Information Systems, The University of Melbourne, Carlton, VIC 3052, Australia. \\This article has been accepted for publication in \textit{Elsevier Future Generation Computer Systems: Special Collection on Advances in Quantum Computing: Methods, Algorithms, and Systems.}}
\ead{sumanraj@uchicago.edu}
\author[inst3,inst4]{Siva Sai}
\ead{siva.sai@nus.edu.sg}
\author[inst2]{Yogesh Simmhan}
\ead{simmhan@iisc.ac.in}
\author[inst1]{Kyle Chard}
\ead{chard@uchicago.edu}
\author[inst4]{Rajkumar Buyya}
\ead{rbuyya@unimelb.edu.au}

\affiliation[inst1]{organization={Department of Computer Science, University of Chicago},
            addressline={5730 S Ellis Ave}, 
            city={Chicago},
            state={IL},
            country={USA}}

\affiliation[inst2]{organization={Department of Computational and Data Sciences, Indian Institute of Science},
            addressline={CV Raman Road}, 
            city={Bengaluru},
            state={KA},
            country={India}}

\affiliation[inst3]{organization={Department of Electrical and Computer Engineering, National University of Singapore},
            addressline={Engineering Drive 3}, 
            country={Singapore}}     
            
\affiliation[inst4]{organization={Quantum Cloud Computing and Distributed Systems (qCLOUDS) Lab, School of Computing and Information Systems, The University of Melbourne},
            addressline={Carlton}, 
            city={Melbourne},
            state={VIC},
            country={Australia}}            

\begin{abstract}
    High-performance computing (HPC) has evolved over decades through multiple architectural transitions, from vector supercomputers to massively parallel CPU clusters and GPU-accelerated systems, continuously expanding the frontier of scientific discovery. With the emergence of quantum processing units (QPUs) as practical computational accelerators, a new opportunity arises to further extend this trajectory by integrating quantum and classical computing paradigms. Building on the emerging vision of Quantum Integrated High-Performance Computing (QHPC), this paper contributes a full-stack layered architecture that integrates QPUs as first-class accelerators within, not merely alongside, the classical HPC software and hardware stack, with tight, on-premise quantum-classical coupling as its defining characteristic. The architecture comprises of unified resource management, quantum-aware scheduling, hybrid workflow orchestration, middleware and programming abstraction, interconnect technologies, and a tiered execution model enabling seamless workload partitioning across classical and quantum backends. A central aspect of this architecture is a strong user requests abstraction layer that exposes heterogeneous resources through a unified job submission interface, similar in spirit to existing schedulers such as Slurm, allowing users to describe workloads in a consistent template independent of underlying compute type or location. Drawing insights from prior accelerator integration eras, we outline how QHPC can support emerging workloads in quantum chemistry, materials discovery, combinatorial optimization, and climate modeling. We conclude by highlighting open challenges in building scalable, reliable, and programmable quantum-classical infrastructures that seamlessly connect global users to heterogeneous compute resources for future quantum-classical HPC ecosystems. 
\end{abstract}

\begin{keyword}
    Quantum Computing \sep High-Performance Computing \sep Classical-Quantum Interfaces \sep Foundations \sep Co-Scheduling
\end{keyword}

\end{frontmatter}

\section{Introduction }
High-Performance Computing (HPC) has been the principal engine of scientific computing for over five decades, enabling predictive modeling and large-scale data analytics across materials discovery~\cite{chen2024accelerating}, genomics~\cite{wang2024big}, astrophysics~\cite{habib16hacc}, and many other domains. The trajectory of scientific computing has always been defined by its most fundamental unit of progress: the \textit{architectural paradigm shift}. 
Contemporary exascale platforms such as Frontier~\cite{10.1145/3581784.3607089}, Aurora~\cite{stevens2019aurora}, and El Capitan exemplify this paradigm shift in architecture. Yet, even at exascale, irreducibly hard problem classes, such as molecular simulations of biologically and chemically relevant systems at full quantum-mechanical fidelity, remain computationally inaccessible~\cite{cao2019quantum}. It is precisely this wall of classical intractability that motivates integrating quantum computation into the HPC stack. 
We are now at a point in time where Quantum Processing Units (QPUs) are poised to become first-class computational accelerators within HPC ecosystems alongside Graphics Processing Units (GPUs), Field-Programmable Gate Arrays (FPGAs), and Application-Specific Integrated Circuits (ASICs). The resulting paradigm, increasingly referred to in the recent literature as ``\textit{Quantum Integrated High-Performance Computing}'' (QHPC), represents not merely the addition of another heterogeneous accelerator, but a qualitative expansion of the computational state space available to scientific workloads: where GPU nodes scale the numerical throughput of classical algorithms linearly with node count, multi-QPU configurations could exploit entanglement-mediated quantum parallelism to extend the accessible Hilbert space exponentially with qubit count~\cite{10.1145/3007651}.

The heterogeneous accelerator model that is common today did not emerge trivially. The integration of GPUs into HPC required rethinking programming models, memory hierarchy co-design, high-bandwidth interconnects, compiler toolchains, and performance portability abstractions. Similarly, FPGAs have been integrated as reconfigurable accelerators for streaming, low-latency, and bit-level parallel workloads. The unifying lesson is that architectural heterogeneity is sustainable only when (a) the accelerator delivers a demonstrable computational advantage for well-defined workload classes, and (b) the system software stack can seamlessly orchestrate heterogeneous execution. In this context, QPUs represent the next logical accelerator in this trajectory, not as a wholesale replacement for classical HPC, but as a tightly integrated co-processor targeting problem classes with provable or empirically observed quantum advantage.

Major institutional initiatives have reinforced this trajectory of QHPC. The U.S. Department of Energy's Oak Ridge National Laboratory (ORNL) is emerging as a flagship quantum-HPC integration facility, with a US\$~$125$ million budget through $2030$~\cite{ornlQSCRenewal2025}. ORNL is deploying an NVIDIA GB200 NVL72 system alongside IQM and Quantum Brilliance QPUs, creating the world's first testbed for side-by-side benchmarking of GPU-accelerated simulation, QPU execution, and hybrid quantum-classical workflows at leadership-computing scale~\cite{ornl2025quantum}. Japan's RIKEN has demonstrated the largest-scale quantum-HPC co-execution to date: a co-located IBM Quantum Heron processor coupled to all 152,064 classical compute nodes of the Fugaku supercomputer~\cite{ibm2026qcsc}. The Center for Strategic and International Studies (CSIS) identified in March 2026 that integrating quantum computers into U.S. supercomputers is a strategic imperative, while noting that the U.S. currently lags Europe and Japan in the number of deployed hybrid quantum-supercomputing systems~\cite{csis2026quantum}. Europe has similarly committed that Germany and France now host Pasqal neutral-atom QPUs within their HPC facilities~\cite{pasqal2026vision}.

To the best of our knowledge, this work offers one of the early visions of a layered QHPC architecture for next-generation hybrid computing infrastructure that integrates QPUs as first-class accelerators \emph{within}, rather than merely alongside, the classical HPC software and hardware stack, tightly coupling on-premises QPUs with conventional CPUs and GPUs under a unified workflow and resource management framework. We use tight coupling in a specific, deployment-anchored sense: it refers to \textit{on-premise co-location} of the QPU within the HPC facility, such that quantum and classical resources share the facility's internal low-latency interconnect fabric rather than being reached across a wide-area network (WAN). This vision aims to bridge the gap between quantum and classical paradigms by providing an integrated execution model that maximizes hardware utilization, minimizes data transfer overhead, and abstracts complexity through intelligent scheduling and orchestration(\S~\ref{sec:architecture}). The paper is conceptual and architectural in scope, and we do not include original experiments or benchmarks; we have identified empirical validation on emerging quantum platforms as a future direction. The rest of the paper is organized as follows: We provide a background of HPC evolution (\S~\ref{sec:background}), lessons learned from CPU--GPU integration (\S~\ref{sec:lessons-learnt}), and a detailed review of the existing Quantum-HPC systems (\S~\ref{sec:existing-systems}). In \S~\ref{sec:embrace-quantum}, we discuss the recent breakthroughs in quantum technologies along with fundamentals of quantum computing. We highlight a range of applications that stand to benefit from these integrated quantum-classical infrastructures (\S~\ref{sec:applications}), and discuss key challenges that arise in such QHPC systems, many of which are likely to shape research directions over the next decade (\S~\ref{sec:challenges}), and finally conclude the paper in \S~\ref{sec:conclusions}.

\section{Background } \label{sec:background}
In this section, we briefly discuss the history and emergence of High-Performance Computing (HPC).

\begin{figure*}[t]
    \centering
    \includegraphics[trim={4.5cm 0 4.5cm 0}, clip, width=0.85\linewidth, height=6cm]{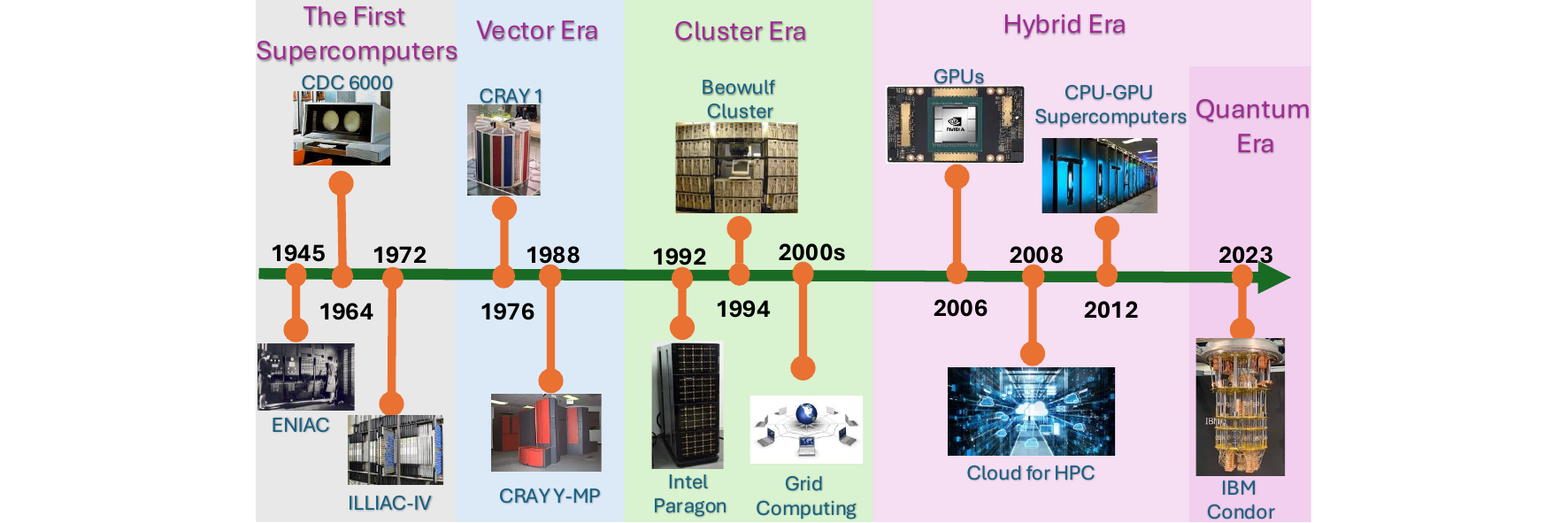}
    \caption{Timeline of the emergence of hybrid HPC and Quantum era}
    \label{fig:timeline}
\end{figure*}

\subsection{Early Supercomputers and the Vector Era: 1940s -- 1990}

HPC began with early digital computers such as the ENIAC (1945)~\cite{goldstine1946electronic}, built for wartime scientific computation. Control Data Corporation's CDC~6000~\cite{Swensen2011} (1964) was the first machine tagged a ``true supercomputer'' (500 KFLOPS--1 MFLOPS), and the ILLIAC-IV~\cite{1450577} pioneered parallelism with a 256-element 64-bit processing array. From the mid-1970s, vector processing and shared-memory multiprocessing dominated: Seymour Cray's Cray~1~\cite{russell1978cray} (1976) reached 133 MFLOPS, followed by the Cray X-MP~\cite{19822} (1982) and Cray Y-MP (1988), the latter scaling to eight vector processors at 333 MFLOPS each.

\subsection{The Cluster Era: 1990 -- 2010} 
While vector processing worked well with 4 or 8 processors, memory contention prevented a 64 or 128-processor machine from working efficiently. This barrier required a transition from shared memory to distributed memory, where each processor has a dedicated memory space. The Message Passing Interface (MPI) standard~\cite{snir1998mpi,walker1996mpi} (1994) was introduced as a unified programming model for distributed systems. The 1990s brought massively parallel processors (MPPs) like Intel's
Paragon (1992), Thinking Machines Corporation's CM-1 (SIMD) to CM-5 (MIMD) and IBM’s SP2. These systems interconnected thousands of processors, enabling scalability for large problems and offering a processing speed of up to 59.7 GFLOPS.
Further, the \textit{Beowulf cluster movement} democratized HPC by enabling researchers to build clusters using commodity hardware and open-source software (typically Linux)~\cite{sterling2002beowulf}. \textit{Grid computing} later emerged to link such clusters globally~\cite{foster2003grid}.

\subsection{GPU and Hybrid Era: 2010 -- Present}

In the 2000s, HPC shifted from increasing clock speeds to scaling through \textit{multi-core architectures}, soon complemented by general-purpose \textit{Graphics Processing Units (GPUs)} offering massive parallelism and improved energy efficiency. Programming frameworks such as CUDA~\cite{nickolls2008cuda} (2006) and OpenCL~\cite{munshi2009opencl} (2009) made these accelerators practical for scientific simulation, deep learning, and large-scale data analytics. Modern HPC has entered the \textit{exascale era}: the \textit{Frontier} supercomputer at Oak Ridge National Laboratory (ORNL)~\cite{10.1145/3581784.3607089} was the first to exceed $1.38\times10^{18}$ FLOPS, and \textit{El Capitan} at Lawrence Livermore National Laboratory (LLNL)~\cite{elcapitan} overtook it in 2024 at $1.7$~exaFLOPS. Both exemplify \textit{heterogeneous hybrid architectures} pairing multi-core CPUs with high-performance GPUs. El Capitan, for instance, combines over $11$ million cores from $43,808$ AMD EPYC CPUs and $43,808$ AMD Instinct MI300A GPUs.

\noindent Overall, the evolution of HPC over the years embodies the relentless pursuit of higher performance, scalability, and efficiency. As HPC approaches the limits of classical architectures, the focus is gradually shifting toward integrating quantum technologies as the next frontier in computation.

\section{Embracing the Quantum Era}\label{sec:embrace-quantum}

The advent of the quantum era signifies a paradigm shift in computational science, leveraging the principles of quantum mechanics to address classes of problems that are computationally infeasible for even the most advanced classical high-performance computing systems. By definition, ``\textit{A quantum computer is a computational device that exploits the principles of quantum mechanics, such as superposition, entanglement, decoherence and interference, to process information using quantum bits (qubits), enabling the solution of certain problems exponentially faster than classical computers}.'' Analogous to Central Processing Unit (CPU) for classical computers, we have Quantum Processing Unit (QPU) as the brain of a quantum computer~\cite{merritt2022whatisaqpu}. A list of the most popular and emerging QPUs have been shared in Table.~\ref{tab:emerging_qpus}, and the anatomy of a QPU has been discussed in \S~\ref{sec:appendix:qpu}.
Next, we briefly discuss the fundamental mechanisms underlying quantum computation.

\begin{table*}[t]
\centering
\tiny
\caption{Types of Qubits (as of October 2025)}
\setlength{\tabcolsep}{1pt}
\begin{tabular}{l|c|l|l}
\toprule
\textbf{Qubit Type} & \textbf{Typical Gate Speed} & \textbf{Medium of Information} & \textbf{Example System}\\ \hline\hline
Superconductivity & $10-100$ns & Josephson junctions and microwave resonators & IBM Eagle, Google Sycamore \\ 
Trapped-Ion & $1-100 \mu$s & Hyperfine states of ions in electromagnetic traps & IonQ Aria, Honeywell H1 \\ 
Photonics & <1 ns & Polarization or path of photons & Xanadu Aurora\\
Neutral-atom & $1-100 \mu$s & Rydberg states of laser-cooled neutral atoms & QuEra Aquila, Atom Computing \\ 
Quantum Dots (Spin) & $10-100$ns & Electron spin in quantum dots or silicon structures & Intel Horse Ridge, TU Delft Spin Q\\
Topological & $\approx1\mu$s~\textit{(Theoretical)} & Majorana zero modes in superconducting nanowires & Microsoft Majorana 1\\
NV-Center & $10-100$ns & Electron spin in nitrogen-vacancy centers in diamond & IBM Quantum Diamond Lab\\
Fluxonium & $100ns-1\mu$s & Superconducting loops with Josephson junction arrays & MIT Lincoln Lab Fluxonium Device\\
Silicon Donor & $1-10\mu$s & Spin states of phosphorus donors in silicon & UNSW SiMOS Qubits\\
NMR & $10-100\mu$s & Nuclear spin states in molecules under magnetic fields & Liquid-state NMR Systems at IBM \& Oxford  \\ 
Electron-on-Helium & $1-10 \mu$s & Electron charge confined on the surface of helium & Experimental academic setups \\ 
CQED & $10-100$ns & Trapped atoms in high-finesse cavities & Yale QLab, ENS Paris\\ 
\bottomrule
\end{tabular}
\label{tab:qubit-technologies}
\end{table*}

\subsection{Principles of Quantum Computing}

\noindent \textit{Superposition:} 
A qubit (quantum bit) is the basic unit of quantum information, analogous to the classical bit. The major qubit technologies are summarized in Table.~\ref{tab:qubit-technologies}. Unlike a classical bit that exists only in state 0 or 1, a qubit can exist in a coherent superposition of both states~\cite{nielsen2010quantum}.
Mathematically, a qubit in superposition is represented as:
    $|\psi\rangle = \alpha |0\rangle + \beta |1\rangle$,
where $\alpha$ and $\beta$ are complex numbers called probability amplitudes satisfying the normalization condition $|\alpha|^2 + |\beta|^2 = 1$. Superposition enables quantum computers to evaluate multiple computational paths in parallel. For example, a system of $n$ qubits can represent $2^n$ possible states simultaneously, providing an exponential increase in the effective state space compared to classical bits.\\

\noindent \textit{Entanglement: }
Entanglement is a unique quantum correlation between qubits, where the state of one qubit cannot be described independently of the state of another, even if they are spatially separated. The simplest illustration of entanglement is given by the \textit{Bell states}. For example, two qubits in the entangled state, $|\Phi^+\rangle={\frac {1}{\sqrt {2}}}(|00\rangle +|11\rangle )$ form an \textit{equal superposition} of the basis states $|00\rangle$ and $|11\rangle$, each occurring with a probability of $(1/\sqrt{2})^2 = 0.5$. Measuring one qubit instantaneously determines the state of the other.\\

\noindent \textit{Decoherence: }
Decoherence refers to the process by which a quantum system loses its coherence and transitions from a quantum superposition to a classical state. This collapse of the quantum state can occur intentionally through measurement or unintentionally due to interactions with the external environment, such as thermal fluctuations or electromagnetic noise. When a qubit in superposition is measured, it collapses to one of its basis states, $|0\rangle$ or $|1\rangle$, with probabilities $|\alpha|^2$ and $|\beta|^2$, respectively.\\

\noindent \textit{Interference: }
Quantum interference arises from the wave-like nature of quantum particles. During computation, multiple computational paths interfere with each other: correct paths are amplified (constructive interference) while incorrect paths cancel out (destructive interference). This mechanism underlies the efficiency of algorithms such as \textit{Grover’s search}~\cite{grover1996fast}, which amplifies correct solutions, and \textit{Shor’s algorithm}~\cite{shor1999polynomial}, which exploits interference to reveal periodic structures.

\begin{table}[t]
\centering
\footnotesize
\caption{Emerging Quantum Processing Units (QPUs) as of November 2025}
\setlength{\tabcolsep}{1pt}
\begin{tabular}{l|l|c|c|l}
\hline
\textbf{Organization} & \textbf{QPU} & \textbf{Launched In} & \textbf{Qubits} & \textbf{Qubit Technology} \\ \hline\hline
Pasqal & Orion Beta & Nov 2025 & 100 & Neutral-Atom \\
Quantinuum & Helios & Nov 2025 & 98 & Trapped-Ion\\
Caltech & - & Sept 2025 & 6,100 & Neutral-Atom \\ 
Quantum Brilliance & Quoll & Sept 2025 & 6 & NV-Center (Diamond)\\
IQM & Radiance & June 2025 & 150 & Superconducting \\
D-Wave Quantum & Advantage 2 & May 2025 & $\sim$4,400 & Quantum Annealing \\
Fujitsu, RIKEN & - & April 2025 &  256 & Superconducting \\
IonQ & Tempo & March 2025 & 64 & Trapped Ion \\ 
Microsoft & Majorana-1 & Feb 2025 & 8 & Topological Qubits \\
USTC & Zuchongzhi 3.0 & Dec 2024 & 105 & Superconducting \\
Google & Willow & Dec 2024 & 105 & Superconducting \\
IBM & Heron (R2) & Nov 2024 & 156 & Superconducting \\ 
CAS (China) & Xiaohong & Jan 2024 & 504 & Superconducting \\ 
IBM & Condor & Dec 2023 & 1,121 & Superconducting \\ 
Atom Computing & - & Oct 2023 & 1,180 & Neutral-atoms \\ 
Intel & Tunnel Falls & June 2023 & 12 & Spin \\
QuEra & Aquila & Nov 2022 & 256 & Neutral Atom \\ 
Xanadu & Borealis & June 2022 & 216 & Photonics \\
\hline
\end{tabular}
\label{tab:emerging_qpus}
\end{table}

\subsection{State-of-the-Art Quantum Computing Architectures}

State-of-the-art quantum computing architectures demonstrate these trends: (i) fault-tolerant error-correcting substrates (notably the surface-code family) and the hardware layouts that realize them, (ii) layered (modular) design patterns that separate physical control from logical layers to manage vastly different time-scales and responsibilities, and (iii) new low-overhead code families and modular network topologies that promise dramatic reductions in physical-qubit overhead. Representative examples include the silicon donor surface-code architecture that maps surface-code error correction onto a planar donor lattice with shared-control gate arrays~\cite{doi:10.1126/sciadv.1500707}, and the layered architecture framework that prescribes distinct physical, error-correction, logical, and application layers -- enabling independent optimization of hardware and compilation stacks~\cite{jones2012layered}. 

Recent advances move beyond surface codes such as advanced compilation-aware layout methods minimize surface-code resource overhead by exploiting long-range Bell-state preparation and edge-disjoint path routing~\cite{beverland2022edpc}; modular ``quantum-system-on-chip'' (QSoC) hardware concepts show how tiled chips and entanglement multiplexing can scale physical qubit counts while simplifying control wiring and connectivity~\cite{li2024heterogeneous,riera2024modular}; bosonic mode encodings offer hardware-efficient logical qubits and new optical error-correction protocols that trade fewer modes for improved lifetime and loss resilience~\cite{albert2022bosonic,hastrup2022cat}; and very recent LDPC-based, ``bicycle'' (bivariate bicycle) architectures propose modular high-rate quantum LDPC codes with explicit logical instruction sets and compilation strategies that can lower overhead by an order of magnitude relative to surface-code architectures for many workloads~\cite{yoder2025tourgrossmodularquantum}. 
Together, these developments point to a hybrid engineering roadmap: optimize device-level layouts for native fault tolerance, adopt layered and modular designs for composability and control, and explore bosonic and qLDPC encodings to substantially reduce the physical-qubit cost of large-scale, fault-tolerant quantum computation.

\section{Lessons Learnt from GPU and Hybrid Era}\label{sec:lessons-learnt}

In this section, we discuss how the transition from traditional CPUs to GPU-accelerated and hybrid architectures has shaped contemporary approaches to parallelism, scalability, and programming models, offering valuable insights for the emerging quantum computing landscape.

\subsection{Heterogeneity as a Design Principle}
Modern HPC systems no longer rely on a single, uniform processor type; instead, heterogeneity has become a central design principle. Sophisticated designs now couple CPUs with GPUs, but extend further to domain-specific accelerators (TPUs, FPGAs, NPUs, ASICs, etc.), unified memory architectures, and hierarchical scheduling for mixed workloads~\cite{10.1145/2788396,fusco2024understandingdatamovementtightly}. This principle provides three key benefits: (1) \textit{Cost-effective specialization}: higher performance and energy efficiency; (2) \textit{Resource multiplexing}: matching tasks to the most suitable accelerator; and (3) \textit{Scalability}: enabling more flexible and efficient heterogeneous systems. Studies show that even within a single GPU, heterogeneity (e.g., multi-instance GPUs, virtualized partitions) and task scheduling across heterogeneous processing elements can yield up to $1.87\times$ throughput gains~\cite{saroliya2023hierarchical}.

This architectural evolution toward fine-grained and hierarchical heterogeneity in classical HPC systems offers valuable insights for emerging quantum--classical platforms, where managing the interplay between QPUs and conventional accelerators will be essential to achieve scalable hybrid computing.

\subsection{Task Parallelism and Workload Decomposition}
GPUs are designed with thousands of small cores that can execute many tasks concurrently, making them ideal for computations that can be divided and run at the same time. These have a significant impact on how compute-bound kernels are expressed and scheduled, and improve performance by orders of magnitude for DNN workloads~\cite{nvidia2020a100}. Task-based runtimes, such as StarPU let programmers express computation as a directed acyclic graph (DAG) of tasks with data dependencies~\cite{10.1007/978-3-642-03869-3_80}. These runtimes have been deployed in HPC environments, including linear algebra, dense and sparse solvers, and various data analytics pipelines, to achieve near-optimal task-to-resource mapping under varying load conditions. Further, runtimes like TimeGraph offer a real-time GPU scheduler at the device-driver level to provide prioritization and isolation of GPU workloads in multitasking environments, overcoming the limitations of non-preemptive GPU execution~\cite{kato2011timegraph}.
This lesson translates directly to quantum systems, where algorithmic parallelism (e.g., in superposition) must be explicitly harnessed through optimized circuit decompositions.

\subsection{Software Ecosystem and Programmability}
The past two decades have proven that hardware performance alone is insufficient and robust software ecosystems and programming models are equally critical. This is true for GPU accelerators, where CUDA, OpenCL and its companion libraries made GPUs accessible and productive~\cite{owens2007survey}, and is already shaping the trajectory of quantum computing (where SDKs such as Qiskit, Cirq and PennyLane form the bridge between algorithms and QPUs).  
Modern ML frameworks, notably PyTorch~\cite{paszke2019pytorch} and TensorFlow~\cite{10.5555/3026877.3026899}, provide high-level libraries that simplify GPU programming by providing abstractions that manage the underlying parallel computations. This enables researchers to scale from a single GPU to multi-node, multi-GPU clusters with relatively modest engineering effort~\cite{nguyen2019machine}. 

This lesson has direct applications in quantum computing. Just like the software ecosystems unlocked the full potential of GPUs, quantum SDKs and frameworks are essential to harness the capabilities of QPUs and enable scalable quantum-classical applications.

\subsection{Scalability through Abstraction and Modularity}
HPC has long demonstrated that scalability requires both modular hardware design and software abstractions. For instance, the Modular Supercomputing Architecture (MSA) developed within the DEEP project series integrates different compute modules with specific performance characteristics, supporting a wide range of applications from computationally intensive simulations to data-intensive artificial intelligence workflows~\cite{riedel2021practice,suarez2019modular}. 
Software abstraction layers such as MPI (Message Passing Interface) for inter-node communication~\cite{walker1996mpi}, OpenMP for intra-node parallelism~\cite{gabriel2004open}, and container orchestration (e.g., Singularity~\cite{kurtzer2017singularity}, Docker~\cite{10.5555/2600239.2600241}) allow applications to scale without requiring developers to manage low-level hardware details. Recent performance analysis on the Frontier and El Capitan supercomputers shows that MPI combined with GPU-aware scheduling can improve throughput and reduce communication overhead in large-scale scientific simulations~\cite{wilfong2025testing}. Furthermore, NVIDIA introduced Multi-Instance GPU (MIG), which enables a single physical GPU to be partitioned into multiple isolated instances, each with dedicated compute cores and memory, enabling concurrent workloads without interference~\cite{li2022miso}. 

This experience with modular, heterogeneous CPU--GPU systems and software abstractions provides a roadmap for building scalable, hybrid quantum--classical platforms.

\subsection{Co-Design of Hardware and Algorithms}
The GPU and hybrid era reinforced the importance of hardware--software co-design for performance portability. 
Large-scale machine-learning systems illustrate this co-design in practice. Megatron-LM~\cite{10.1145/3458817.3476209,shoeybi2020megatronlmtrainingmultibillionparameter} showed that careful intra-layer model parallelism combined with optimized communication patterns sustains high scaling efficiency across hundreds of GPUs when training multi-billion parameter transformers. Complementary frameworks, such as Microsoft DeepSpeed~\cite{rasley2020deepspeed} and ZeRO~\cite{rajbhandari2020zero}, demonstrate that memory-centric decompositions enable training trillion-parameter models by alleviating per-device memory pressure and reducing redundant communication. 
Architectures like the NVIDIA A100/H100, with exposed Tensor Cores and CUDA-X libraries, amplify these benefits, showing that performance and energy efficiency are fully realized only when algorithms and compilers are co-
designed with hardware features. 

This lesson is now underpinning quantum algorithm design, such as matching circuit depth, qubit connectivity, and gate sequences, to specific QPU topologies.

\section{QHPC System Architecture}\label{sec:architecture}

\begin{figure}[t]
    \centering
    \includegraphics[width=0.75\textwidth]{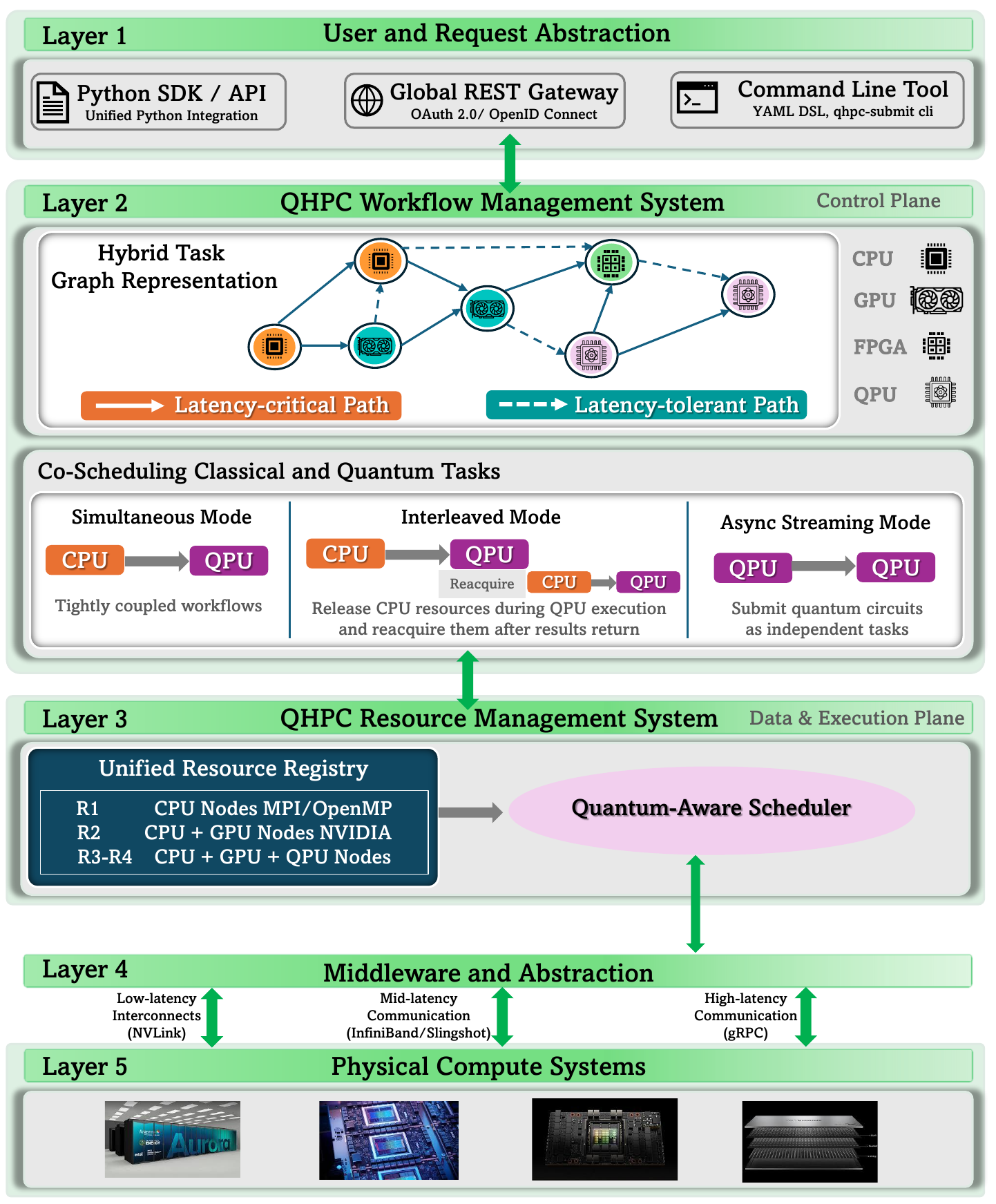}
    \caption{Quantum-Integrated HPC Architecture Layers }
    \label{fig:vision}
\end{figure}

In this section, we propose the QHPC architecture that constitutes a full-stack, heterogeneous computing architecture that treats QPU as a first-class accelerator within, not alongside, the classical HPC software and hardware stack. Tight coupling is realized by the R3 tier, which is the defining characteristic of the proposed architecture, owing to its low latency. Loose coupling, by contrast, is off-premise, cloud/remote access over WAN, realized by the R4 tier as a graceful-degradation fallback for latency-tolerant workloads.
Our overarching expectation is that a user, from any institution or country, can submit a workload through a single, technology-agnostic entry point, and the QHPC system decompose, route, schedule, and execute that workload across the available combination of CPUs, GPUs, FPGAs, specialized AI accelerators, and QPUs, using the mechanisms specified in the layers below. Figure~\ref{fig:vision} illustrates the QHPC architecture across five architectural layers: the User and Request Layer, the Workflow Management Layer, the Resource Management Layer, the Middleware and Abstraction Layer, and the Physical Compute Layer. 

\subsection{Design Philosophy: The Spectrum Between Slurm and Kubernetes}

Before detailing the individual layers, a more fundamental architectural question must be addressed: what \textit{execution model} appropriately captures hybrid quantum--HPC workloads? Classical HPC has largely converged on a batch-processing paradigm, using cluster managers and job schedulers like Slurm, where jobs are queued, resources are allocated exclusively, and execution is deterministic, time-bounded, and aware of system topology. In contrast, cloud-native systems adopt a microservice and container-orchestration model (e.g., Kubernetes), where long-lived services expose APIs, resources are elastically provisioned, and workloads are packaged for portability and resilience.

Hybrid quantum--classical workflows do not fit cleanly into either paradigm. Algorithms such as VQE and QAOA are inherently iterative and stochastic, coupling quantum circuit execution with classical optimization loops whose convergence behavior depends on measurement noise and shot statistics. At the same time, they require strict, topology-aware allocation of QPU resources to ensure reproducibility and isolation, which is incompatible with fully elastic service models. Consequently, hybrid workloads occupy an intermediate regime that challenges both batch-oriented determinism and service-oriented elasticity, motivating the need for a new execution abstraction tailored specifically to QHPC systems. This also anticipates the direction in which major vendors (IBM, NVIDIA, HPE) are converging their quantum-HPC integration blueprints.

\begin{figure}
    \centering
    \includegraphics[width=0.5\linewidth]{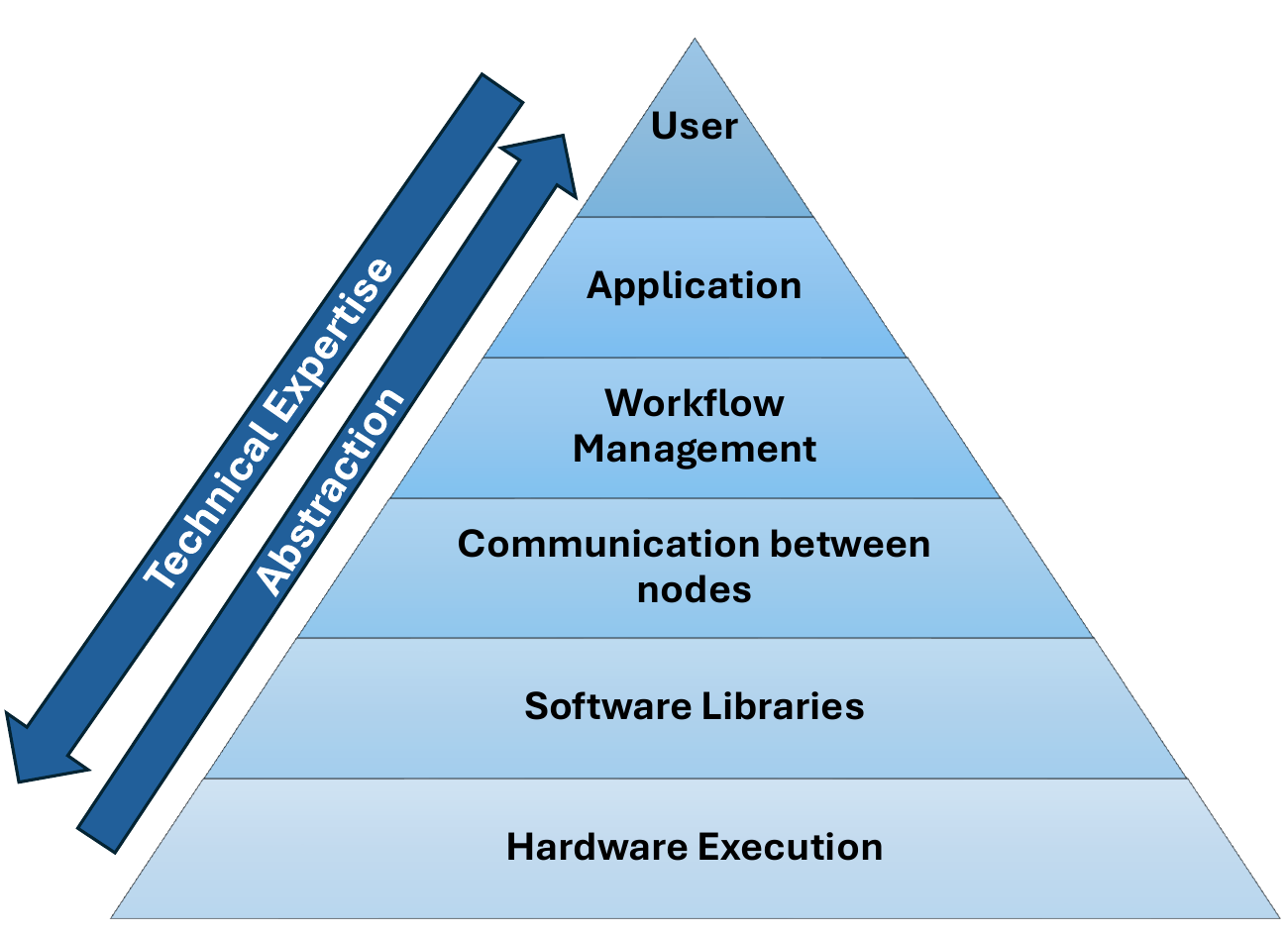}
    \caption{Abstraction vs Expertise for QHPC Systems}
    \label{fig:hierarchy}
\end{figure}

\subsection{Layer 1: User and Request Abstraction}\label{sec:layer1}
The entry point to the QHPC system is a \textit{unified submission interface} that is agnostic to the underlying compute modality. Users submit a Hybrid Workload Descriptor (HWD), a structured specification analogous to a Slurm or PBS job script but extended to include quantum circuit requirements, via one of three access channels. First, a command-line tool (\texttt{qhpc-submit}) compatible with \texttt{sbatch} that accepts HWDs in YAML or a domain-specific language and transparently forwards classical workloads to the HPC scheduler while routing quantum-enabled jobs through the QHPC workflow manager; second, a Python SDK exposing a \texttt{QHPCJob} abstraction for programmatic integration within Python, C++, or Fortran codes, unifying interactions with frameworks such as Qiskit primitives and OpenMP-Q-style quantum offloading; and third, a Kubernetes-fronted REST gateway secured via OAuth 2.0/OpenID Connect that enables global, allocation-free access for job submission, monitoring, telemetry, and result retrieval. Fig.~\ref{fig:hierarchy} shows that higher abstraction makes systems easier to use and requires less expert knowledge.

The HWD itself comprises two logical components. The \textit{classical logic descriptor} specifies traditional HPC requirements, including CPU/GPU counts, memory footprint, accelerator tier, MPI topology, data I/O paths, and execution time constraints. The \textit{quantum logic descriptor} defines the quantum workload, including circuit representation (OpenQASM 3.0, QIR, or CUDA-Q kernels), qubit count and connectivity constraints, shot budget or target statistical confidence, supported QPU modalities (superconducting, trapped-ion, neutral-atom, or best-available), and fallback execution policies such as GPU-based emulation via cuQuantum or tensor-network simulation.

\subsection{Layer 2: QHPC Workflow Management System (WMS)}\label{sec:layer2}
The QHPC Workflow Management System is the \textit{control plane} of the entire architecture. Its responsibilities are decomposing hybrid workloads into executable task graphs, managing inter-task data dependencies, scheduling tasks to the appropriate compute tier, and monitoring execution progress.

\subsubsection{Hybrid Task Graph Representation}\label{sec:layer2-1}
The WMS models a hybrid workload as a Directed Cyclic Task Graph (DCTG), extending classical DAG-based workflow model to support the iterative feedback loops inherent in variational algorithms, in which quantum measurement results are fed back into classical optimization nodes. Each node is typed as CPU (control flow and preprocessing), GPU (vectorized and ML workloads), QPU (quantum circuit execution producing probabilistic outputs on hardware or emulators), or FPGA (ultra-low-latency control tasks such as pulse generation and real-time QEC decoding that are unsuitable for GPU execution). The QHPC architecture is inherently extensible to new accelerator modalities, as any device exposed via a job scheduler plugin or Kubernetes Device Plugin can be incorporated into its unified resource hierarchy and scheduled natively.

\subsubsection{Dependency Analysis and Critical Path Scheduling}\label{sec:layer2-2}
Upon receiving a hybrid workload, the WMS performs \textit{semantic dependency analysis} that goes beyond the structural data dependencies captured in the DCTG. It identifies \textit{latency-critical paths}, chains of $CPU\to QPU\to CPU$ dependencies in which the classical side is idle while awaiting QPU results, and \textit{latency-tolerant paths}, sequences of independent QPU circuit evaluations that can be batched and submitted in parallel to exploit QPU multi-programming or multiple-QPU parallelism. For latency-critical paths, the WMS routes the workload to co-located QPU and CPU resources connected via a low-latency fabric, targeting sub-millisecond round-trip times. For latency-tolerant paths, the WMS can distribute circuit evaluations across multiple QPUs at different geographic sites.

\subsubsection{Co-Scheduling Classical and Quantum Tasks}\label{sec:layer2-3}
The co-scheduling problem of allocating classical HPC nodes and QPU access is the central scheduling challenge of QHPC. The QHPC WMS tackles the co-scheduling of classical HPC nodes and QPU access through three per-job modes. \textit{Simultaneous mode} co-allocates CPUs and QPUs for the full job enabling tightly coupled workflows like VQE and QAOA where classical and quantum steps alternate frequently.
\textit{Interleaved mode} releases CPU resources during QPU execution and reacquires them after results return, reducing idle time but adding scheduling overhead; the system dynamically switches to simultaneous mode when quantum phases are too short to benefit from interleaving.
\textit{Asynchronous streaming mode} submits quantum circuits as independent tasks while the classical pipeline continues execution, making it suitable for loosely coupled workloads such as quantum machine learning and batched kernel evaluation.

\subsubsection{Analytical Model of the Co-Scheduling Modes}\label{sec:cosched-model}
To ground the mode-selection policy quantitatively rather than qualitatively, we model a hybrid variational job (e.g., VQE or QAOA) of $N$ iterations. Each iteration comprises a classical phase of duration $t_c$ (parameter update and pre-/post-processing) and a quantum phase of duration $t_q$ (circuit submission, QPU execution, and the classical--quantum round trip). Let $s$ denote the overhead of releasing and reacquiring classical resources around a quantum phase (queue re-entry and context setup). We compare the three modes on the job wall-clock $T$ and the idle classical resource-time $I$ (allocation reserved but not computing).

\emph{Simultaneous mode} co-allocates classical and quantum resources for the whole job:
\begin{equation}
T_{\text{sim}} = N\,(t_c + t_q), \qquad I_{\text{sim}} = N\,t_q,
\end{equation}
so the classical allocation sits idle for a fraction $t_q/(t_c+t_q)$ of the job. \emph{Interleaved mode} releases classical resources during each quantum phase and reacquires them afterward:
\begin{equation}
T_{\text{int}} = N\,(t_c + t_q + s), \qquad I_{\text{int}} \approx 0,
\end{equation}
eliminating idle reservation, but adding $N\,s$ to the wall-clock while freeing $N\,t_q$ of classical resource-time for other jobs. \emph{Asynchronous streaming mode} applies when the $M$ circuit evaluations of a phase are mutually independent, giving a wall-clock bounded by $\max\!\big(N\,t_c,\ (M/p)\,t_q\big)$ for QPU parallelism $p$, rather than the sum.

\textit{Mode selection.} Interleaving is beneficial only when the classical resource-time it frees exceeds the overhead it introduces, i.e., the per-iteration net benefit
\begin{equation}
B = t_q - s > 0 \iff t_q > s .
\end{equation}
The workflow manager therefore selects interleaved mode when $t_q > s$ and reverts to simultaneous mode when quantum phases are too short ($t_q \le s$) to amortize the switch overhead. This threshold also explains the tier dependence: for tightly coupled R3 execution (small $t_q$) simultaneous mode dominates, whereas for remote R4 access (large $t_q$ from WAN latency and cloud queues) interleaving pays off. As an illustration, with $t_c = 50$~ms and switch overhead $s = 200$~ms, an R3 quantum phase of $t_q = 2$~ms yields $t_q \ll s$, so simultaneous mode is preferred (idle fraction $\approx 4\%$); an R4 quantum phase of $t_q = 4$~s yields $t_q \gg s$, so interleaving frees $\approx 99\%$ of the classical allocation at $\approx 5\%$ wall-clock cost.

\textit{Scheduler cost.} For $n$ queued jobs and $k$ candidate QPUs, the quantum-aware scheduler applies the feasibility filter in $\mathcal{O}(nk)$ and ranks feasible QPUs by their Quantum Suitability Score in $\mathcal{O}(k \log k)$ per job, giving $\mathcal{O}(nk \log k)$ per scheduling cycle, which is polynomial and dominated by the resource-inventory size.

\subsection{Layer 3: QHPC Resource Management System (RMS)}\label{sec:layer3}
The QHPC Resource Management System is the \textit{data and execution plane}, managing the inventory, allocation, scheduling, and runtime monitoring of all physical compute resources.

\subsubsection{Unified Resource Model}\label{sec:layer3-1}
The RMS maintains a \textit{Unified Resource Registry} (URR) that lists every available compute resource as an object, along with its type and capabilities.
QPU resources are modeled as dynamic resources. Unlike GPUs, whose performance is largely fixed, a QPU’s effective capability depends on its current calibration state, which can drift frequently. The RMS periodically polls vendor calibration APIs (e.g., every 15 minutes) and updates the resource profile with metrics such as gate fidelity and coherence time. This allows the scheduler to make calibration-aware decisions, selecting the most reliable QPU available for a given job.

\subsubsection{Multi-Tier Resource Hierarchy}\label{sec:layer3-2}
The QHPC resource hierarchy is organized into four tiers. \textit{R1} comprises CPU-only nodes, multi-socket systems (e.g., AMD EPYC Genoa) interconnected via high-speed InfiniBand -- serving classical workloads, MPI-parallel simulations, and hosting classical optimizers using MPI+OpenMP. \textit{R2} includes CPU+GPU accelerated nodes (e.g., NVIDIA Blackwell-based systems), supporting GPU-accelerated simulation, deep learning, molecular dynamics, and quantum circuit emulation via cuQuantum or CUDA-Q.
\textit{R3} represents tightly integrated CPU+GPU+QPU nodes, where QPUs are co-located with classical resources and connected via low-latency interconnects (e.g., NVLink), enabling sub-microsecond feedback. \textit{R4} consists of remote or cloud-accessed QPUs (e.g., IBM, IonQ, Amazon Braket), exposed via REST/gRPC APIs with higher latency, and thus suited for loosely coupled workloads. We emphasize that R4 serves as an off-premise access and fallback tier, rather than the tightly coupled execution model that defines QHPC. It supports latency-tolerant workloads and provides access to QPU modalities that are not yet deployable within on-premise HPC facilities, while latency-critical hybrid workloads are executed on the co-located R3 tier.
Across all tiers, the system supports heterogeneous QPU modalities, allowing the scheduler to map workloads to the most suitable hardware.

\subsubsection{The Quantum-Aware Scheduler}\label{sec:layer3-3}

The QHPC scheduler is designed as a layered policy engine that can be implemented on top of existing resource managers (e.g., Slurm) through extensible plugins and interfaces. The \textit{classical scheduling layer} provides standard capabilities such as fair-share policies, backfilling, and accelerator-aware allocation for CPUs and GPUs. The \textit{QPU availability layer} augments this with calibration-aware selection, using a Quantum Suitability Score (QSS) that combines gate fidelity, connectivity compatibility, queue wait time, and access latency to choose the most appropriate QPU; jobs are issued virtual QPU device tokens encapsulating this context.
A \textit{co-scheduling optimizer} handles tightly coupled hybrid jobs by jointly allocating classical resources and QPUs within shared time and latency constraints, while a fallback policy ensures progress under resource scarcity by dynamically routing workloads to GPU-based emulation, queuing for QPU access, or notifying users of degraded execution modes.

\subsection{Layer 4: Middleware and Abstraction Layer}\label{sec:layer4}

The Middleware Layer is the software stack that sits between the WMS/RMS and the physical hardware, providing hardware-agnostic programming abstractions, circuit compilation and transpilation, error mitigation, and communication protocols.

\subsubsection{Quantum Intermediate Representation and the Compilation Pipeline}\label{sec:layer4-1}
All quantum circuits submitted to QHPC are first lowered to the Quantum Intermediate Representation (QIR), an LLVM-based, hardware-agnostic IR that enables portable optimization and backend-specific lowering, analogous to LLVM in classical HPC. The compilation pipeline then proceeds in four stages: (i) logical optimization, including gate cancellation, commutation, and depth reduction; (ii) device mapping and qubit routing, where logical qubits are mapped to hardware with noise-aware placement and SWAP insertion based on current calibration data; (iii) pulse-level optimization (for supported superconducting QPUs), generating device-specific control pulses to minimize execution time; and (iv) error mitigation selection, where strategies such as ZNE, PEC, or CDR are chosen based on circuit characteristics and the target QPU’s noise profile, or omitted for fault-tolerant systems.

\subsubsection{The Classical--Quantum Communication Protocol}\label{sec:layer4-2}
Communication between classical HPC nodes and QPU control systems is handled through a \textit{three-layer Quantum Communication Protocol Stack}. The \textit{transport layer} uses low-latency interconnects such as NVLink, PCIe Gen~5, or CXL for co-located (R3) systems, achieving round-trip latencies of $<4,\mu s$ (controller--GPU) and $<10,\mu s$ (QPU--HPC), while remote (R4) QPUs are accessed via HTTPS/gRPC over Ethernet or InfiniBand. The \textit{session layer} maintains persistent connections, abstracts vendor-specific APIs, and handles authentication, job submission, and device lifecycle events. The \textit{data serialization layer} encodes shot-based outputs into efficient formats (e.g., NumPy arrays) and transfers them via a lightweight protocol Quantum Result Transport Protocol (QRTP), optimized for high-throughput communication of $10^4$--$10^6$ samples per circuit evaluation.

\subsubsection{Quantum-Aware Programming Interface}\label{sec:layer4-3}
The middleware exposes a \textit{Quantum-Aware Programming Interface }(QAPI) with three abstraction levels. At the highest level, users access prebuilt hybrid algorithms (VQE, QAOA, QSVM, QAE) that take domain objects such as Hamiltonians or datasets and internally handle circuit generation, compilation, QPU execution, error mitigation, and post-processing, targeted at domain scientists without quantum programming expertise. The circuit level supports direct submission via OpenQASM 3.0, CUDA-Q, Qiskit, or PennyLane for users requiring fine-grained control over quantum workflows. At the lowest level, pulse APIs provide direct hardware control for \textit{R3} systems with pulse access, enabling calibration and device-level experimentation.

\subsection{Layer 5: Physical Compute Layer}\label{sec:layer5}

The physical layer of QHPC consists of compute hardware and a three-tier interconnect fabric. \textit{Interconnect hierarchy.} Intra-node NVQLink provides sub-4 µs latency and $\sim$64 Gb/s bandwidth for tight QPU--GPU coupling and real-time QEC, while inter-node InfiniBand NDR or Slingshot-11 supports MPI-scale communication. A wide-area quantum-federation overlay enables secure access to remote QPUs (R4) over WAN (10--100 ms latency), suitable only for latency-tolerant workloads. Superconducting QPUs require $\sim$15 mK operation with 5--30 kW cryogenic overhead and strict isolation (magnetic shielding, vibration control), leading to physically separated QPU pods connected via FPGA-based control electronics to the HPC fabric. In contrast, photonic and NV-center QPUs can be rack-integrated with standard CPU/GPU systems.\\

\noindent Overall, the proposed QHPC architecture is designed to evolve gracefully: as QPU qubit counts increase, error rates decrease, and coherence times lengthen along published vendor roadmaps (IBM targeting 10,000+ two-qubit gates by 2028; Google targeting fault-tolerant logical qubits by 2029), the R4-tier workloads of today, those currently requiring remote access or GPU-emulation fallback, will progressively migrate to tightly coupled, on-premise R3 physical QPU execution. As a consequence, R3 tier will host increasingly complex applications that today are only tractable classically, and R4 path will recede over time.\\

\noindent The feasibility of realizing these mechanisms is supported by our comparison of existing systems in Table~\ref{tab:qhpc-comparison}: each capability dimension (D1--D7) is already provided, in full or in part, by one or more of the implemented or demonstrated systems surveyed there. This indicates that the individual building blocks of the architecture already exist in practice, and that the design can be instantiated by composing and extending these established components rather than by introducing untested primitives.


\section{Existing Quantum Integrated HPC Systems}\label{sec:existing-systems}

It is essential to comprehend the current landscape of existing QHPC systems, both for integrating and improving upon them in future work. In this section, we discuss existing QHPC systems and their features.

Mohseni et al.~\cite{mohseni2026buildquantumsupercomputerscaling} present a comprehensive roadmap for scaling quantum computers from hundreds to millions of qubits into what they term quantum supercomputers. The work spans the full stack, discussing qubit hardware and modalities, cryogenic infrastructure, interconnects, quantum error correction, compilation, and the classical co-processing and HPC integration required to operate quantum processors at scale. It establishes the long-term hardware and systems trajectory that motivates tight quantum--classical integration. Compared to our work, they do not specify a unified submission interface, co-scheduling modes, or a calibration-aware resource manager. Similarly, Beck et al.~\cite{BECK202411} offer a perspective on integrating quantum computing resources into scientific HPC ecosystems. The paper examines integration models spanning on-premise and cloud deployments, the software-stack components needed for programming, compilation, scheduling, and resource management, and representative scientific use cases, and it articulates the principal challenges and a roadmap for embedding QPUs within HPC facilities.

 Mantha et al. \cite{mantha2025pilot} introduced Pilot-Quantum, a QHPC middleware for task, resource, and workload management. Pilot-Quantum is built for quantum applications that are based on task parallelism, such as circuit cutting systems (used to divide and execute large quantum circuits) and hybrid algorithms (like variational approaches). Pilot-Quantum is built on the basis of a rigorous analysis of existing application execution patterns and quantum middleware systems. It also supports high-level programming frameworks such as Pennylane and Qiskit. The authors demonstrated the capabilities of the system through quantum machine learning, circuit cutting, and mini-apps (simple and representative kernels focused on performance bottlenecks) scenarios. 

Shehata et al. \cite{shehata2025bridging} proposed a hardware-agnostic QHPC software to support both futuristic fault-tolerant quantum computers and the current noisy-scale quantum devices, thus presenting a comprehensive software stack. The proposed QHPC software includes standardized APIs for resource management, a quantum gateway interface, and a scheduling mechanism to handle both interleaved and parallel quantum-classical workloads. A few important elements of the framework include - a quantum programming interface to abstract hardware details; a quantum platform manager to integrate various quantum hardware systems; a resource management system to efficiently coordinate quantum and classical resources; and a tool chain for quantum circuit execution and optimization. Similar to Pilot-Quantum, the authors demonstrated the proposed framework with the implementation of hybrid quantum-classical algorithms like variational quantum linear solver, concluding the framework's ability to handle complex workflows while considering resource utilization maximization.  

Burgholzer et al. \cite{burgholzer2025munich} introduce the Munich Quantum Software Stack (MQSS), a modular, multi-layered architecture for tightly coupled hybrid workflows that integrates QPUs as first-class HPC accelerators. Its core ``middle-end'' layer comprises a Quantum Resource Manager \& Compiler Infrastructure (QRM \& CI) built on a Multi-Level Intermediate Representation (MLIR), enabling hardware-aware, dynamic compilation and optimization driven by live system telemetry. This layer includes a bi-level scheduler that coordinates with classical HPC managers (e.g., Slurm) to orchestrate hybrid jobs and minimize idle time across classical and quantum resources. The ecosystem exposes high-level applications through front-end adapters and interfaces heterogeneous hardware via the Quantum Device Management Interface (QDMI), which abstracts device control and data acquisition protocols.

Guerreschi et al. \cite{guerreschi2020intel} presented the Intel Quantum Simulator (IQS), a large-scale simulator capable of full state-vector simulation for quantum circuits up to 42 qubits by using HPC infrastructure. The architecture uses OpenMP to employ a hybrid parallelization strategy for shared memory and the Message Passing Interface (MPI) for distributed memory. At the MPI, the quantum states are partitioned across process nodes, and operations on local qubits are executed without the need for inter-process communications (the global qubits require data exchange between paired processes). The authors implement a \enquote{pool} functionality that divides global HPC resources into independent groups, which allows for parallel simulation of different quantum states. The proposed architecture enables cloud-ready deployment and accelerates specific workloads like stochastic noise modeling (implemented via ensemble averaging and noise gates) and variational algorithms like Quantum Approximate Optimization Algorithm (QAOA) driven by classical Particle Swarm Optimization. IQS includes a Python wrapper that is implemented with the help of Pybind11, allowing it to work as an HPC backend for popular quantum software frameworks, including Cirq, Qiskit, and ProjectQ. It also includes Dockerfiles that can be built into Docker containers to facilitate easier deployment on HPC clusters and cloud computing platforms. IQS provides methods for direct computation of exact probabilities and expectation values from the state vector without collapsing the quantum state. This is in contrast to the physical quantum hardware that requires multiple shots to estimate outcomes. The proposed methodology was benchmarked on the SuperMUC-NG supercomputer.

With the goal of enabling large-scale simulations of quantum circuits with the help of HPC systems, Wang et al. \cite{wang2023enabling} developed NWQ-Sim (Northwest Quantum Simulator), a high-performance quantum circuit simulator. The authors integrated NWQ-Sim with XACC, a programming framework for hybrid quantum-classical applications, and executed variational quantum eigensolver and quantum phase estimation for quantum chemistry problems at a large scale. NWQ-Sim includes particular optimizations to handle the large computational overhead of VQE simulations, such as caching and reusing post-ansatz states, gate fusion, and direct expectation value calculation. The authors demonstrate the potential of optimized simulators and HPC resources to advance quantum chemistry. For instance, a proposed component of the system - caching reduced the gates required for energy evaluation by 3 to 5 orders of magnitude.  

Esposito et al. \cite{esposito2023hybrid} proposed a strategy for the orchestration of hybrid quantum-classical workloads in QHPC  systems. The proposed method interleaves classical and quantum tasks, making use of heterogeneous job launches. It operates under a multiple program multiple data (MPMD) paradigm to manage distinct quantum and classical tasks in parallel. The authors employ the message passing interface (MPI) for data exchange between components, and the Classiq software suite and Qiskit Aer simulator, respectively, for quantum circuit synthesis and quantum circuit simulation. As a part of their case study, the authors worked with the Harrow-Hassidim-Lloyd (HHL) quantum algorithm, which is used to solve linear systems of equations that are derived from the discretization of linear differential equations. They show that the extraction of state vectors and circuit synthesis are the most time-consuming processes and that communication overhead via MPI was negligible. 

Schusler et al. \cite{schusler2023towards} demonstrated the integration of quantum and HPC servers in a real-world setup. The authors integrated the Dutch national HPC Center (SURF) with the Quantum Inspire platform with the help of a distributed infrastructure consisting of two distinct Slurm clusters (C1 and C2). The HPC center acts as the main host (C1), which submits JSON-formatted batch jobs to the second cluster (C2) through a REST API. The second cluster is co-located with the QUantum Inspire API. The proposed system utilizes a fast task manager (dispatcher) to orchestrate interactions between the co-located classical runtime and the quantum runtime. A request/reply pattern is employed via ZeroMQ to minimize the latency to the microsecond scale. This allows the system to execute Python-based hybrid tasks iteratively, wherein the classical runtime continues to generate circuits for the quantum runtime until a stop condition is met. 

Shehata et al. \cite{shehata2024framework} proposed a quantum framework (named QFw) that is designed to streamline the \enquote{loose integration} of quantum simulation resources with on-premise HPC environments, and validated QFw on the Frontier supercomputer. This methodology improves over the traditional per-process simulator model and utilizes Slurm's heterogeneous job feature to parallelly allocate distinct sets of compute nodes for the classical application, and a dynamic simulation environment for the quantum part. The proposed architecture orchestrates these resources with the help of a Distributed Execution Framework (DEFw) that employs Remote Procedure Calls (RPC) to manage communication between the simulation backends and the classical application. The workflow is managed by two important components - the quantum task manager (QTM) and quantum platform manager (QPM). The framework uses the PMIx runtime environment to execute the simulations to establish a distributed virtual machine. Chundury et al. \cite{chundury2024qfw} extended QFw by introducing a lighteright Python library to allow multiple front ends' interaction with QFw and by integrating it with Northwest Quantum Simulator (NWQ-Sim). 

Caldwell et al. \cite{caldwell2025platform} proposed a framework, NVQLINK, designed to tightly couple HPC resources with a quantum system controller to support latency-critical workloads like quantum error correction. The core methodology used a scalable \enquote{real-time interconnect} built on commodity Ethernet and remote direct memory access (RDMA) technology. 
The authors defined four distinct time domains - physical, deterministic, application, and real-time to orchestrate this heterogeneous system and extended the Cuda-Q model with new abstractions, allowing the developers to use multi-level intermediate representation (MLIR) to compile unified C++ programs that address FPGAs, CPUs, and GPUs, within a single logic flow. 
 The proposed framework implements distinct compilation workflows based on the latency sensitivity. The compiler enforces an ahead-of-time strategy for systems with high latency sensitivity, where the complete ISA programs are pre-compiled, executed, and uploaded to FPGAs. On the other hand, the framework supports just-in-time compilation and interactive execution or low-latency sensitivity systems. It achieved sub-4 microsecond round-trip latencies in a proof-of-concept setup.

Chen et al. \cite{chen2024quantum} presented a middleware scheme designed for HPC to enable distributed quantum computing through a hybrid workflow. The proposed QHPC framework integrates classical distributed scheduling with multi-GPU acceleration to manage the computational intensity of quantum simulations. In the proposed framework, a variational quantum eigensolver is employed to approximate ground states efficiently. This is followed by a classical preprocessing stage where Convolutional Neural Networks are used for feature extraction on quantum data, which is subsequently re-encoded into a final quantum circuit layer for scalar classification output generation. The framework uses a middleware stack incorporating the NVIDIA cuQuantum SDK and Pennylane to address the issue of computational intensity of the simulations. 

Zhan et al. \cite{zhan2025full} proposed a modular, hardware-agnostic framework that is designed to integrate quantum computing into existing HPC environments with three core extensions. Firstly, a quantum interface library connects traditional HPC languages (C, C++, Fortran) with Python quantum SDKs via a hybrid MPI client-server execution model. This separates quantum circuit synthesis and execution from classical pre-/post-processing, which allows for asynchronous parallelism managed by the Slurm workload manager, with the goal of minimizing QPU idle time. Secondly, the Adaptive Circuit Knitting (ACK) hypervisor works as a workload distributor that utilizes tensor network-based analysis for partitioning large quantum circuits into smaller ones, thus enabling execution on smaller-scale, distributed QPUs. Finally, the framework also incorporates a quantum compiler extension that ingests LLVM IR and Quantum IR (QIR) to generate hardware-retargetable binaries, with a goal of optimizing compile times and enabling frontend-agnostic execution. 

Slysz et al. \cite{slysz2025hybrid} present a hybrid classical--quantum HPC environment at the Poznań Supercomputing and Networking Center integrating two room-temperature ORCA Computing PT-1 photonic QPUs into a standard data-center infrastructure via Ethernet. These QPUs are coupled with NVIDIA V100 and H100 GPUs to support heterogeneous workloads, with an extended Slurm scheduler managing CPUs, GPUs, and QPUs in a multi-user setting. The software stack leverages NVIDIA CUDA-Q and exposes photonic QPUs as HTTP REST services, enabling execution of hybrid variational algorithms such as binary bosonic solvers and quantum neural networks. These workflows iteratively offload quantum subroutines while performing classical post-processing on HPC nodes. Elsharqawy et al. \cite{elsharkawy2024integration} proposed a unified quantum platform to standardize the low-level integration of different quantum hardware modalities into HPC ecosystems using a co-designed hardware and software architecture. At the software level, the proposed methodology utilizes a unified runtime library that translates platform-agnostic quantum intermediate representation code into a unified quantum instruction set architecture. The translated instruction set architecture is capable of handling both quantum and classical control flow operations, and they are offloaded to a generalized quantum control processor (QCP) via shared memory. The QCP microarchitecture extends previous designs to be technology-agnostic at the hardware layer with he help of modular hardware accelerators like units for sorting and atom detection, that are dynamically activated to support specific physical quantum platforms like neutral atoms or superconducting circuits. 

Beyond on-premise HPC--QPU integration, several adjacent research lines inform our design. In quantum resource management and scheduling, modeling and simulation toolkits such as iQuantum for quantum computing environments~\cite{https://doi.org/10.1002/spe.3331} and fidelity-aware orchestration~\cite{nguyen2025qfor} complement the pilot- and Slurm-based approaches discussed above. In quantum cloud and service-oriented computing, the quantum-serverless paradigm and the QFaaS framework~\cite{nguyen2025quantum}, together with the choreography of quantum--classical FaaS workflows across hybrid clouds~\cite{jha2025ccgrid} and broader quantum-cloud perspectives~\cite{GILL20261}, expose QPUs as on-demand services and target loosely coupled, cross-cloud execution. These directions are complementary to our architecture: they correspond to the remote, service-based access captured by our R4 tier, whereas our contribution focuses on unifying such access with tightly coupled, on-premise execution.\\

\begin{table*}[t]
\centering
\scriptsize
\setlength{\tabcolsep}{5pt}
\renewcommand{\arraystretch}{1.3}
\caption{Taxonomy of dimensions used to compare quantum-integrated HPC systems.}
\label{tab:qhpc-taxonomy}
\begin{tabular}{p{3.6cm} p{4.5cm} p{7.6cm}}
\toprule
\textbf{Dimension} & \textbf{Description} & \textbf{Representative categories ($\rightarrow$: increasing capability)} \\
\midrule
Interface Abstraction (D1) & Interface through which users express and submit hybrid workloads. & SDK/framework-level $\rightarrow$ middleware API $\rightarrow$ unified technology-agnostic interface. \\
Coupling regime (D2) & Physical proximity between classical and quantum resources. & Loose (remote/cloud over WAN) $\rightarrow$ tight (on-premise, co-located, low-latency) $\rightarrow$ spectrum spanning both. \\
Co-scheduling model (D3) & Strategy for jointly allocating classical and quantum tasks. & None/static offload $\rightarrow$ single fixed mode $\rightarrow$ multiple, dynamically selected modes. \\
Resource management (D4) & How are compute resources modeled, monitored, and selected. & Static specifications $\rightarrow$ calibration-aware (dynamic QPU state) $\rightarrow$ calibration-aware and multi-tier hierarchy. \\
QPU-modality support (D5) & Breadth of quantum hardware technologies the system can target. & Single modality $\rightarrow$ a few specific modalities $\rightarrow$ modality-agnostic (heterogeneous). \\
Architectural scope (D6) & Range of system layers addressed. & Single component/layer $\rightarrow$ middleware layer $\rightarrow$ full-stack. \\
Maturity  (D7) & Level of realization and supporting evidence. & Conceptual/architectural $\rightarrow$ prototype/proof-of-concept $\rightarrow$ implemented, demonstrated, or deployed. \\
\bottomrule
\end{tabular}
\end{table*}

\begin{table*}[t]
\centering
\scriptsize
\setlength{\tabcolsep}{1pt}
\renewcommand{\arraystretch}{1.3}
\caption{Comparison of existing quantum-integrated HPC systems with the proposed QHPC architecture across the seven dimensions defined in Table~\ref{tab:qhpc-taxonomy}. $\checkmark$ = supported (rightmost category), $\bullet$ = partial (middle category), \ding{55} = not addressed (leftmost category). }
\label{tab:qhpc-comparison}
\begin{tabular}{p{3cm} p{7cm} ccccccc}
\toprule
\textbf{System} & \textbf{Primary focus / scope} & \textbf{D1} & \textbf{D2} & \textbf{D3} & \textbf{D4} & \textbf{D5} & \textbf{D6} & \textbf{D7} \\
\midrule
Mohseni et al.~\cite{mohseni2026buildquantumsupercomputerscaling} & Vision: scaling quantum supercomputers & \ding{55} & $\bullet$ & \ding{55} & \ding{55} & $\bullet$ & \ding{51} & Conceptual \\
Beck et al.~\cite{BECK202411} & Perspective: integrating QC into HPC ecosystems & $\bullet$ & $\bullet$ & $\bullet$ & \ding{55} & $\bullet$ & \ding{51} & Conceptual \\
Pilot-Quantum~\cite{mantha2025pilot} & Task/resource/workload middleware & $\bullet$ & \ding{55} & $\bullet$ & \ding{55} & $\bullet$ & $\bullet$ & Implemented \\
Shehata et al.~\cite{shehata2025bridging} & HPC--quantum convergence stack & \ding{51} & $\bullet$ & $\bullet$ & \ding{55} & \ding{51} & $\bullet$ & Design + demo  \\
MQSS~\cite{burgholzer2025munich} & Modular, tightly coupled SW stack & \ding{51} & $\bullet$ & $\bullet$ & $\bullet$ & \ding{51} & \ding{51} & Implemented \\
IQS~\cite{guerreschi2020intel}, NWQ-Sim~\cite{wang2023enabling} & Large-scale circuit simulation on HPC & \ding{55} & \ding{55} & \ding{55} & \ding{55} & \ding{55} & \ding{55} & Benchmarked \\
Esposito et al.~\cite{esposito2023hybrid} & MPMD hybrid orchestration & $\bullet$ & $\bullet$ & $\bullet$ & \ding{55} & \ding{55} & $\bullet$ & Demo \\
Sch\"usler et al.~\cite{schusler2023towards} & HPC--QPU integration demonstrator & $\bullet$ & $\bullet$ & \ding{55} & \ding{55} & \ding{55} & $\bullet$ & Real-world demo \\
QFw~\cite{shehata2024framework,chundury2024qfw} & Quantum framework (loose integration) & $\bullet$ & \ding{55} & $\bullet$ & \ding{55} & $\bullet$ & $\bullet$ & Validated \\
NVQLINK~\cite{caldwell2025platform} & Tight HPC--controller coupling (QEC) & $\bullet$ & $\bullet$ & $\bullet$ & \ding{55} & $\bullet$ & $\bullet$ & Proof-of-concept\\
Chen et al.~\cite{chen2024quantum} & GPU-accelerated hybrid middleware & \ding{55} & $\bullet$ & \ding{55} & \ding{55} & \ding{55} & $\bullet$ & Demonstrated \\
Zhan et al.~\cite{zhan2025full} & Full-stack hardware-agnostic framework & \ding{51} & $\bullet$ & $\bullet$ & \ding{55} & \ding{51} & \ding{51} & Framework\\
Slysz et al.~\cite{slysz2025hybrid} & Multi-user photonic hybrid environment & $\bullet$ & $\bullet$ & \ding{55} & \ding{55} & $\bullet$ & $\bullet$ & Real deployment \\
Elsharkawy et al.~\cite{elsharkawy2024integration} & Unified low-level HW/SW platform & \ding{51} & $\bullet$ & \ding{55} & \ding{55} & \ding{51} & $\bullet$ & Architectural \\
\midrule
\textbf{Ours} & Full-stack unified architecture & \ding{51} & \ding{51} & \ding{51} & \ding{51} & \ding{51} & \ding{51} & Architectural \\
\bottomrule
\end{tabular}
\end{table*}

\noindent \textit{\uline{Synthesis and comparison with our architecture:}}
Table~\ref{tab:qhpc-taxonomy} defines the seven dimensions (D1--D7) along which the discussed QHPC systems can be characterized, and Table~\ref{tab:qhpc-comparison} positions the surveyed works and our proposed architecture against them. The comparison shows that existing efforts each address a subset of the QHPC problem, most commonly a single coupling regime (D2), a compiler or middleware layer (D6), or a validated point deployment (D7), whereas our architecture is the only entry that spans the full coupling spectrum (D2), provides three dynamically selected co-scheduling modes (D3), and combines calibration-aware selection with an explicit multi-tier resource hierarchy (D4), while also supplying a unified technology-agnostic submission abstraction (D1), heterogeneous modality support (D5), and full-stack scope (D6). In particular, dynamic multi-mode co-scheduling (D3) and calibration-aware tiered resource management (D4) are the dimensions least addressed in prior work. Where our design shares a capability with prior work, for example, hardware-agnostic interfaces (D1, D5) with \cite{burgholzer2025munich,zhan2025full,elsharkawy2024integration}, or full-stack scope (D6) with the vision papers\cite{mohseni2026buildquantumsupercomputerscaling,BECK202411}, we regard those systems as complementary rather than competing. We also note the corresponding limitation: unlike the implemented or demonstrated systems in the table, our architecture is presented at the conceptual and architectural level (D7), with empirical validation identified as future work. Existing quantum-HPC proposals focus on individual capabilities. Our work unifies these capabilities within a technology-agnostic, full-stack architecture while introducing mechanisms for unified workload specification, dynamic co-scheduling, and calibration-aware resource management.

\section{Applications}\label{sec:applications}
In this section, we discuss various applications from multiple domains that can benefit from our proposed QHPC framework (\S\ref{sec:architecture}). 
We survey application domains across which this framework delivers value, ranging from classical HPC workloads already transformed by GPU acceleration to emerging quantum-native and hybrid quantum-classical use cases that can exploit QPUs as they mature.  

\subsection{Computational Chemistry, Drug Discovery, Materials and Molecular Simulation}

In computational chemistry, the case for QPU integration is strongest on theoretical grounds. Classical HPC has advanced molecular simulation through density functional theory (DFT), coupled-cluster methods, and molecular dynamics, yet all of these face an exponential barrier: exact full configuration interaction (FCI) scales combinatorially with system size, rendering strongly correlated molecules classically intractable \cite{mcardleQuantumComputationalChemistry2020}. QPU-accelerated algorithms such as VQE and QPE offer a natural alternative, encoding fermionic wavefunctions directly in Hilbert space and circumventing the exponential memory overhead of classical representations \cite{peruzzo2014variational}, \cite{liHybridQuantumPipeline2024}. 
Similarly, in materials science, the HPC community has pursued classical-scale simulations for decades using density functional theory codes such as VASP, Quantum ESPRESSO, and CP2K. Here again, the fundamental limitation is the same as in computational chemistry: DFT exhibits cubic scaling with system size, restricting exact treatment to systems of fewer than a few thousand atoms, while critical phenomena in strongly correlated materials involve electronic correlations that DFT fails to capture reliably \cite{kimMachineLearningEnergy2026}.

The practical significance of QHPC capability for drug discovery is also substantial. To illustrate how the proposed architecture (\S~\ref{sec:architecture}) would support such a workload end-to-end, we trace it through the five layers. For instance, large-scale screening tasks evaluating millions of candidate ligand conformations for a drug-protein binding calculation can be offloaded to GPU-accelerated classical nodes leveraging ML force fields and docking engines. In our architecture, a user expresses this workload as a single Hybrid Workload Descriptor (\S~\ref{sec:layer1}), which the Workflow Management System (\S~\ref{sec:layer2}) compiles into a Directed Cyclic Task Graph; this screening stage is a latency-tolerant GPU task scheduled on R2 nodes (\S~\ref{sec:layer3-2}). However, a reduced subset requiring chemically accurate binding free energies is selectively escalated to QPU-accelerated VQE or sample-based quantum diagonalization, supported by HPC-driven error mitigation \cite{santagatiDrugDesign2024}. Within the architecture, this escalation is a latency-critical CPU$\rightarrow$QPU$\rightarrow$CPU loop: the Quantum-Aware Scheduler (\S~\ref{sec:layer3-3}) selects a QPU using its Quantum Suitability Score and co-allocates classical and quantum resources in simultaneous mode (\S~\ref{sec:layer2-3}) on a co-located R3 node, while the middleware (\S~\ref{sec:layer4}) lowers the circuit to QIR, applies noise-aware routing and error mitigation, and returns shot statistics to close the variational loop; should no suitable QPU be available, the fallback policy routes the stage to GPU-based emulation on R2. This traces the application through all five layers and shows how the architectural choices support the workload, while remaining a design-level illustration rather than an implemented result. Further, Li et al.~\cite{liHybridQuantumPipeline2024} have also demonstrated such a hybrid quantum computing pipeline for real-world drug discovery, targeting high-fidelity Gibbs free energy estimation for prodrug activation. \\

The energy storage sector presents the most near-term commercial motivation. Hybrid quantum-classical simulations of lithium battery electrolytes, including VQE-based calculations of ground and excited states for salts, have been demonstrated as early proof-of-concept QPU workloads in the context of the IBM quantum-HPC co-design program~\cite{qeomElectrolytes2025}. The U.S. Department of Energy's ARPA-E Quantum Computing for Computational Chemistry (QC3) program has specifically identified quantum simulation of energy materials as priority use cases for the hybrid HPC-QPU paradigm \cite{arpaeQC32024,harnessQuantumEnergyMaterials2026}.

\subsection{Genomics, Life Sciences and Bioinformatics}

Genomics and life sciences represent one of the fastest-growing HPC application domains, driven by the continued decline in sequencing costs and the resulting exponential growth in genomic data volumes. Genome sequencing analysis pipelines are routinely executed on HPC clusters and increasingly on GPU-accelerated nodes, using frameworks such as NVIDIA Parabricks. Within the QHPC framework, classical bioinformatics pipelines continue to be served efficiently on the CPU and GPU tiers, while QPU-based quantum machine learning opens new capabilities for several technically challenging sub-problems \cite{wangGenomicsHPC2024}. Quantum machine learning classifiers, particularly quantum support vector machines (QSVMs) and quantum neural networks, have been demonstrated for genomic variant interpretation, multi-omics integration, and drug-target interaction prediction, tasks that are characterized by high-dimensional feature spaces and limited labeled training data, conditions that favor the quantum kernel advantage \cite{qmlReview2025}. Quantum annealing and QAOA have been applied to protein folding optimization as a QUBO problem, and variational quantum algorithms have been used to predict RNA secondary structure folding energies with higher accuracy than classical dynamic programming approaches on small instances. FreeQuantum is a hybrid quantum-classical pipeline for biomolecular free energy calculations, which in QHPC could be dynamically partitioned between QPUs for quantum-accelerated steps and classical HPC for sampling and ML tasks~\cite{gunther2025use}.

\subsection{Quantum-Accelerated Artificial Intelligence and Machine Learning}

AI workloads constitute the fastest-growing share of HPC center allocations globally. The QHPC framework, whose GPU-accelerated classical tiers already serve these workloads efficiently, additionally opens the possibility of QPU-assisted acceleration for specific AI sub-problems where quantum methods offer a theoretical advantage. Quantum machine learning (QML) encompasses a family of algorithms in which quantum circuits replace or augment classical neural network components. AI methods improve quantum hardware calibration, circuit compilation, and error correction, while quantum hardware accelerates AI training by exploiting quantum kernel methods, quantum sampling, and quantum feature maps that can represent exponentially large feature spaces with polynomial quantum resources \cite{qmlReview2025}, \cite{aiForQuantumNC2025,qmlComprehensive2025}. Reinforcement learning presents a particularly compelling hybrid use case: during training of a quantum-enhanced RL agent for robotic control, a quantum critic network executed on a QPU can process the high-dimensional state representations produced by the environment, while the classical actor network, which must execute in real time during inference, runs entirely on CPU nodes with GPU accelerators~\cite{qaiStatusPerspectives2025}.

\subsection{Earth Science, Climate Modeling and Computational Fluid Dynamics}

The integration of QPU accelerators introduces new capabilities for several sub-problems in the Earth system sciences that are computationally intractable at the classical scale. Ho et al. provide a systematic review of quantum computing applications to climate resilience and sustainability challenges, identifying QML models for climate pattern recognition, quantum optimization for renewable energy dispatch, and quantum-enhanced hydrological modeling as near-term use cases \cite{otgonbaatar2023quantum}. 
Quantum optimization algorithms are well-suited to the combinatorial problems embedded in climate-related logistics, such as the optimal placement of renewable energy generation facilities (wind farms, solar installations) subject to grid topology, transmission constraints, and the optimization of carbon capture~\cite{otgonbaatar2023quantum}.

Computational fluid dynamics (CFD) is one of the most demanding HPC workloads, as resolving turbulent flows at engineering scale Reynolds numbers requires multi-scale simulations that consume millions of CPU--GPU core-hours on leadership systems. The Navier-Stokes equations governing atmospheric and oceanic dynamics are the core computational kernel of climate models, and recent theoretical analysis has demonstrated analytic bounds suggesting that quantum algorithms can achieve exponential speedup over classical CFD solvers for turbulent flow simulation, a result with direct implications for cloud and convection parameterization in climate models as well~\cite{liQuantumAdvantageFluidDynamics2024,garciaInspiredCFD2024}. Complementarily, Syamlal et al. implemented and benchmarked a variational quantum CFD solver on real QPU hardware, establishing the feasibility of solving CFD-governing PDEs on near-term quantum processors and outlining a path toward industrial-scale adoption \cite{syamlalVQCFD2024}. A comprehensive review categorizes quantum-CFD approaches into circuit-based algorithms, quantum-inspired tensor networks, and hybrid variational solvers~\cite{malinvernoCFDReview2025}.

\subsection{Financial Modeling and Combinatorial Optimization}
The financial services industry is one of the most computationally intensive non-scientific domains, spending billions of dollars annually on compute resources for risk management, derivative pricing, portfolio construction, and fraud detection~\cite{orusQuantumFinance2025,dimoulkasQuantumFinance2025}. The most theoretically grounded near-term quantum advantage in finance is in Monte Carlo simulation, where quantum amplitude estimation (QAE) achieves quadratic reduction in sample complexity relative to classical Monte Carlo~\cite{stamatopoulosOptionPricing2020}. For computation of Value-at-Risk (VaR), Conditional Value-at-Risk (CVaR), and derivative pricing under correlated stochastic processes, this speedup translates to either higher accuracy at fixed computational cost or equivalent accuracy at substantially reduced runtime, both outcomes of direct commercial value.
Within the QHPC framework, a financial analyst can submit a portfolio-optimization task specifying the asset count, constraints, and accuracy targets. The scheduler dynamically maps small instances to CPU solvers, medium-scale problems to GPU-accelerated heuristics or quantum-inspired methods, and large, highly correlated cases to QAOA or variational circuits on QPUs. This mirrors hybrid quantum--classical deployments in industry, such as the collaboration of JP Morgan Chase with IBM, where quantum amplitude estimation was used for European option pricing on real QPU hardware, illustrating a pathway toward practical financial quantum acceleration \cite{liHybridQuantumPipeline2024}.

Combinatorial and discrete optimization problems, such as vehicle routing, supply chain scheduling, facility location, network flow, and inventory management, constitute the largest category of industrially relevant NP-hard problems. Ahmed et al. present a recent systematic review identifying 20 studies in 2024 alone on quantum applications in supply chain management, spanning dynamic routing, demand forecasting, and disruption-resilient scheduling via quantum annealing and QAOA \cite{ahmedQCSupplyChain2025}. Similarly, the QED-C’s 2024 use-case assessment highlights route optimization, multimodal scheduling, and emissions-aware logistics as high-impact quantum opportunities, emphasizing continuous route optimization to reduce fuel costs and improve profitability~\cite{qedcTransportLogistics2024}. Complementing this, Zaman et al. document the use of QAOA and quantum-inspired evolutionary methods for carbon-neutral supply chains, linking quantum logistics optimization directly to sustainability objectives~\cite{zamanQCSupplyChain2025}.\\

\noindent Additional domains, including cybersecurity, cryptography, astrophysics, high-energy physics, and nuclear simulations, are also potential candidates that can benefit significantly from the QHPC framework. However, a discussion of these areas is deferred to future work.

\section{Open Challenges}\label{sec:challenges}

In this section, we present the key open challenges for QHPC, each framed in relation to the proposed architecture (\S~\ref{sec:architecture}). For every challenge, we identify both the architectural choice it motivates and the component that provides a concrete substrate for investigating it.

\subsection{Heterogeneous Facility Co-Location: Physical and Operational Co-Design}

\noindent \uline{The open problem}: \textit{How can a shared facility be designed and operated to support CPUs and diverse accelerators (GPUs, FPGAs, ASICs, and QPUs) with differing and often conflicting infrastructure requirements?}

Quantum processors impose infrastructure requirements that differ fundamentally and often conflict with those of classical HPC systems. Superconducting QPUs require cryogenic operation at millikelvin temperatures, strict magnetic shielding, and vibration isolation, while other modalities (e.g., trapped-ion, neutral-atom, photonic) demand optical stability, vacuum environments, and electromagnetic control. In contrast, HPC facilities are optimized for high-density CPU/GPU deployments with significant power draw, liquid cooling, and mechanical activity, which can create interference that may degrade quantum coherence. The lack of standardized specifications for integrating QPUs alongside classical accelerators further necessitates bespoke, facility-specific engineering, as highlighted by recent Open Compute Project efforts \cite{ocpQuantumDataCenter2025}.

This challenge raises several research questions: (i) Can unified thermal management architectures be developed to support both cryogenic QPU environments and high-density accelerator nodes without cross-interference, and what are the implications for power efficiency and facility design? (ii) What electromagnetic isolation and zoning strategies are required to preserve QPU fidelity in environments with high-speed interconnects and multi-kilowatt compute nodes? and (iii) Can modular, rack-scale quantum infrastructure be standardized to enable scalable and repeatable deployment within HPC facilities? Within our architecture, these trade-offs are localized to the Physical Compute Layer (\S~\ref{sec:layer5}) and the R3/R4 tiers (\S~\ref{sec:layer3-2}), which make co-location and isolation decisions explicit and thus open to investigation.

\subsection{Quantum-Classical Workflow Orchestration and the Circuit-Cutting Problem}

\noindent \uline{The open problem}: \textit{How should a workflow system decompose and orchestrate hybrid quantum--classical applications across heterogeneous resources, while supporting low-latency feedback, mid-circuit operations, and circuit partitioning when workloads exceed available QPU capacity?}

Hybrid quantum--classical algorithms such as VQE, QAOA, and QSVM are inherently iterative, with bidirectional data dependencies between quantum circuit execution and classical optimization. This breaks the Directed Acyclic Graph (DAG) assumption underlying existing HPC workflow systems (e.g., Pegasus, Nextflow, Parsl), necessitating dynamically reconfigurable workflows \cite{cranganoreHybridWorkflows2024}. The challenge is compounded by mid-circuit measurement and feedforward, where error correction and control logic require sub-microsecond classical response times~\cite{caldwell2025platform}, far below the latency supported by current workflow managers.

Additionally, circuit cutting and knitting, which is the technique of decomposing a circuit too large for a single QPU into subcircuits executable on smaller QPUs, with classical post-processing to reconstruct the full output distribution, introduces exponential classical overhead in the number of cuts \cite{manthaPilotQuantum2025}. 
Recent empirical profiling shows that workload splitting, using circuit cutting for VQE and data-parallel batching for QSVM, can materially improve throughput and reduce quantum resource footprint (e.g., QSVM up to 3x faster and 3x lower qubit-seconds), but it introduces non-trivial classical overheads (e.g., reconstruction/merge costs that can reach tens of minutes in VQE) and accuracy–mitigation trade-offs that must be surfaced to the runtime and user~\cite{khare2023qce}.

Optimal circuit partitioning is a hard combinatorial optimization for which practical approaches rely on heuristics with significant classical overhead~\cite{khare2023qce}. Designing workflow systems that natively support quantum circuit semantics, including feedback loops, low-latency control, and scalable circuit partitioning, remains an open challenge. Complementarily, choreography of quantum-classical workflows across HPC centers demonstrates that packaging these optimizations as reusable orchestration patterns (fan-out, cross-cloud partitioning and asynchronous/adaptive polling) can yield large system-level gains (e.g., up to 53\% faster execution and 80\% lower cost across hybrid clouds), while coping with long and uncertain QPU queues~\cite{jha2025ccgrid}.

Research questions posed here are: (i) Can the standard workflow DAG model be extended to a directed cyclic graph with bounded depth that captures the iterative feedback structure of variational quantum algorithms, and what scheduling theory is applicable to such cyclic graphs with probabilistic edge weights? (ii) What is the fundamental lower bound on circuit knitting overhead for structured application families such as molecular Hamiltonians and QAOA-based MaxCut, and can exploitable problem structure reduce the classical reconstruction cost from exponential to polynomial scaling? (iii) Can a just-in-time (JIT) quantum circuit compiler be developed that dynamically re-partitions circuits at runtime based on evolving QPU calibration data, analogous to JIT compilation in GPU systems, to optimize fidelity and execution efficiency? The Directed Cyclic Task Graph and path analysis of our workflow manager (\S~\ref{sec:layer2-1}-\ref{sec:layer2-2}) are designed to represent exactly this iterative-feedback and partitioning structure, and are the point at which circuit-cutting and low-latency-feedback strategies would be integrated.

\subsection{A Unified Quantum-HPC Programming Model}

\noindent \uline{The open problem}: \textit{How can the decades-long MPI/OpenMP/CUDA ecosystem be extended to support first-class, portable quantum offloading, enabling HPC developers in C++ or Fortran to invoke QPU kernels with the same abstraction used for GPU acceleration?}

The HPC programming model is organized as a layered stack: MPI for distributed memory communication, OpenMP for shared-memory parallelism, and CUDA/HIP/SYCL for accelerator offloading. This ecosystem has matured over three decades and is deeply integrated into scientific computing codes. In contrast, quantum software frameworks remain largely circuit-centric and Python-driven (e.g., Qiskit, Cirq, PennyLane, CUDA-Q), with no native interoperability with MPI communicators, Slurm schedulers, or C/Fortran data layouts \cite{rallis2025interfacingHPCQC}. This mismatch in programming models, memory representations, and execution semantics forms a fundamental barrier to practical QHPC integration~\cite{rallis2025interfacingHPCQC}. 
Early efforts attempt to bridge this gap: OpenMP-Q \cite{openmqOffloading2025} proposes a pragma-based extension to OpenMP that enables quantum task offloading using the familiar \texttt{\#pragma omp target} paradigm, dispatching quantum circuits to QPU backends from within standard C++ HPC application code; CONQURE \cite{conqureFramework2025} provides an open-source co-execution framework achieving VQE runtime reductions through parallelised OpenMP quantum offloading; and the QIR Alliance's LLVM-based quantum intermediate representation \cite{qir2022} provides a portable circuit IR that can in principle be targeted by any HPC compiler. However, these approaches remain incomplete: they do not fully address encoding MPI-distributed data into quantum states, integrating probabilistic measurement outcomes into classical workflows, or enabling performance-portable execution across heterogeneous QPU modalities.

Possible research questions are: (i) Can an MPI--QPU collective communication primitive, analogous to \texttt{MPI\_Allreduce}, be defined to aggregate shot-based measurement statistics across distributed QPU executions? (ii) What compiler analyses and transformations are required to automatically identify quantum-amenable subcomputations in legacy C++/Fortran HPC codes and translate them into QIR-compliant circuits? (iii) Can a portable performance model, analogous to the Roofline model, be developed for such systems to predict optimal classical--quantum workload partitioning given heterogeneous hardware characteristics? The Hybrid Workload Descriptor (\S~\ref{sec:layer1}), the three-level QAPI (\S~\ref{sec:layer4-3}), and the QIR-based compilation pipeline (\S~\ref{sec:layer4-1}) define the interfaces through which such a unified offloading model could be realized.

\subsection{LLM-Assisted and Autonomous Quantum-HPC Code Generation and Workflow Synthesis}

\noindent \uline{The open problem}: \textit{Can large language models (LLMs) be leveraged to automatically synthesize, optimize, and deploy quantum--classical application code, while jointly accounting for hardware availability and target accuracy constraints, thereby lowering the programming barrier for QHPC systems?}

We argue that a primary bottleneck to democratizing access to QHPC is not hardware availability but software expertise~\cite{rallis2025interfacingHPCQC}. Developing correct and efficient hybrid quantum--classical applications requires expertise spanning domain science, classical HPC programming, quantum algorithms, circuit compilation, error mitigation, and scheduler interaction. While LLMs have shown promise in generating syntactically valid quantum circuits in frameworks such as Qiskit and Cirq, translating between quantum SDKs, and suggesting algorithmic refinements, their ability to produce correct, hardware-aware, and performance-optimized hybrid HPC code remains unproven \cite{aiForQuantumNC2025}.

Key open questions include: (i) Can a domain-specific fine-tuned LLM consistently generate hybrid quantum--classical programs that are both algorithmically correct and hardware-adapted, as measured by compilation success, circuit fidelity, and convergence behavior on standard benchmarks? (ii) What formal verification frameworks can certify the correctness of LLM-generated quantum circuits prior to execution, and what is the computational overhead of such verification at NISQ-scale circuit sizes? (iii) Can an autonomous QHPC agent combining LLM-based code synthesis, hardware-aware performance modeling, and closed-loop feedback from QPU execution continuously improve its code generation and scheduling policies without human supervision? The unified, technology-agnostic submission interface and the high-level QAPI (\S~\ref{sec:layer1}, \S~\ref{sec:layer4-3}) provide a natural target representation for LLM-based synthesis: by abstracting away hardware and scheduling details, they reduce the complexity of the code an autonomous agent must generate.

\subsection{Fault Tolerance, Resilience, and Reproducibility in QHPC Systems}

\noindent \uline{The open problem}: \textit{How should a QHPC system be architected to ensure high availability, reproducibility, and graceful degradation under QPU and classical subsystem failures, particularly for tightly coupled hybrid workloads?}

Classical HPC resilience relies on checkpoint/restart, fault-tolerant communication (e.g., MPI), and deterministic error detection mechanisms such as checksums. These approaches do not translate directly to QPU-based systems: quantum states cannot be copied due to the no-cloning theorem~\cite{nielsen2010quantum}, mid-circuit checkpointing collapses the computation via measurement, and quantum outputs are inherently probabilistic, requiring repeated execution to estimate observables. As a result, QPU faults can manifest as silent degradations, for example, calibration drift may bias expectation values without triggering explicit failure signals, leading to incorrect optimization trajectories in hybrid workflows~\cite{nguyen2025qfor}.
In fault-tolerant regimes, the coupling becomes even tighter: quantum error correction requires real-time classical decoding within sub-microsecond latencies, and decoder failures can directly result in logical qubit loss, as demonstrated in recent real-time QEC experiments by IBM~\cite{ibm2025nighthawk}. This creates a distinctive resilience regime in which classical HPC reliability and quantum hardware stability are co-dependent, and where failure in either subsystem can irreversibly compromise the computation. 

Several open research questions arise: (i) Can a quantum job reproducibility protocol be defined that specifies the minimal set of artifacts required to reproduce a quantum--classical result, and what are the associated storage and computational overheads at scale? (ii) What is the theoretical lower bound on classical compute resources required for real-time QEC decoding in a fault-tolerant QPU with $\mathcal{N}$ logical qubits, and how does this requirement scale as $\mathcal{N}$ grows from tens to thousands of logical qubits? (iii) Can a hybrid checkpointing mechanism be designed that preserves the state of variational optimizations (parameters, gradients, and convergence history) to enable seamless recovery after QPU failures without invalidating previously consumed shot budgets? Calibration-aware resource profiling and the scheduler's fallback policy (\S~\ref{sec:layer3-1}, \S~\ref{sec:layer3-3}), the error-mitigation stage of the compilation pipeline (\S~\ref{sec:layer4-1}), and FPGA-based real-time QEC in the physical layer (\S~\ref{sec:layer5}) are the architectural hooks through which resilience and reproducibility mechanisms can be investigated.\\

\section{Conclusions and Future Directions}\label{sec:conclusions}

In this paper, we present Quantum Integrated High- Performance Computing, which offers a unified architectural vision that integrates CPUs, GPUs, and QPUs as first-class resources within a common stack. This enables seamless execution of hybrid workloads while preserving portability and scalability across heterogeneous systems. By extending established HPC paradigms into the quantum era, the framework lays the foundation for a unified compute continuum for future scientific discovery.

Future research directions span both systems and architectural frontiers. An important direction is the energy-efficient and sustainable operation of large-scale hybrid infrastructures, where quantum, classical, and accelerator-rich resources must be co-optimized under power and thermal constraints. Another promising direction is quantum emulation at scale, where GPU-based QPU simulation becomes a first-class HPC workload, enabling development, debugging, and benchmarking of quantum algorithms before hardware maturity. Extending the hardware abstraction layer, we envision deeper integration of FPGAs as real-time quantum system controllers, error-correction decoders operating within microsecond coherence windows, and low-latency hybrid accelerators for tightly coupled classical-quantum routines. Collectively, these directions point toward a future where heterogeneous compute resources operate as a unified, programmable, and distributed compute continuum.

\section*{Acknowledgements}
The authors would like to thank Kartik Singhal for their valuable input on the article. The research visits of Suman Raj and Siva Sai to the University of Melbourne are funded by an Australian Research Council (ARC) Discovery Project.   
Additionally, Suman Raj was supported by Prime Minister's Research Fellowship, Ministry of Education, India. 
Yogesh Simmhan was supported by a grant from the National Quantum Mission, DST, India.

\bibliographystyle{unsrt}
\bibliography{references}

\appendix

\section{Quantum Processing Units (QPUs)}\label{sec:appendix:qpu}

\begin{figure}[h]
    \centering
    \includegraphics[width=0.7\linewidth]{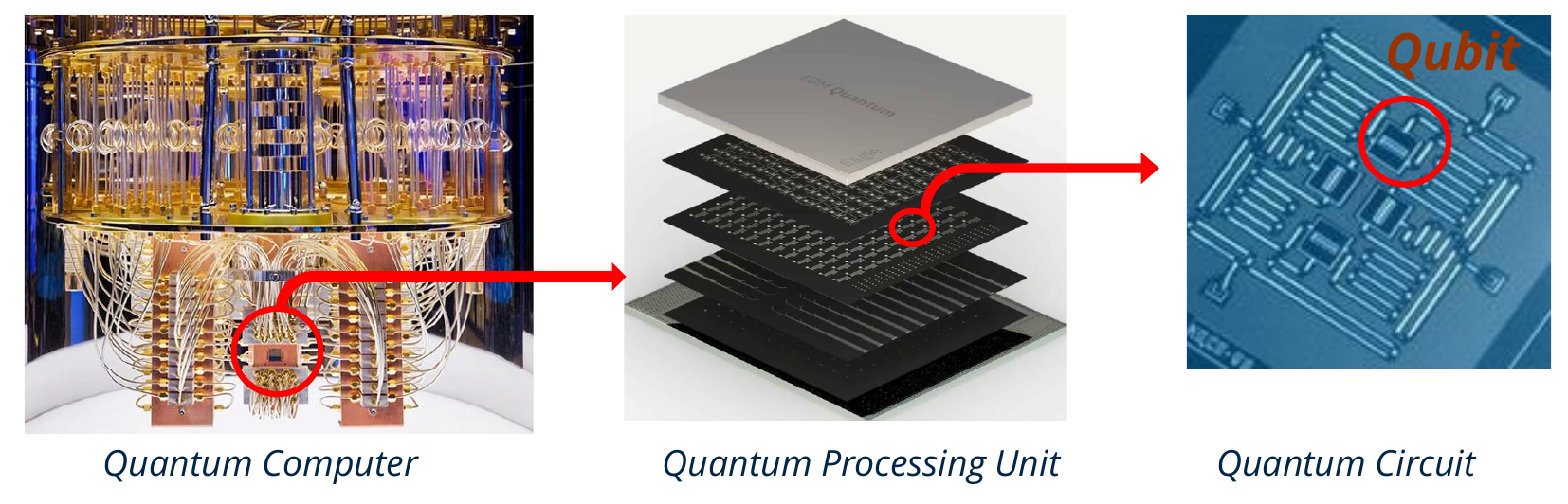}
    \caption{Representative anatomy of a Quantum Computer showcasing a quantum computer, a quantum processing unit, quantum circuit and a qubit. Picture Courtesy: IBM}
    \label{fig:qpu-anatomy}
\end{figure}

The first step to manufacture a QPU is to first realize the physical qubits using one of the technologies discussed in Table.~\ref{tab:qubit-technologies}. We show an illustration in Fig.~\ref{fig:qpu-anatomy} where superconducting qubits are realized as tiny electrical circuits made from superconducting materials such as aluminum or niobium, connected through \textit{Josephson junctions}. These junctions act as non-linear inductors, enabling the circuit to exhibit quantum behavior, i.e., superpositions of current or voltage states. Once individual qubit circuits are designed, they are patterned onto silicon or sapphire substrates using nanofabrication techniques similar to those used in semiconductor manufacturing. Multiple qubits, along with resonators and coupling elements, are integrated onto a single \textit{quantum processor unit} (or chip). This chip is then mounted inside a dilution refrigerator, which cools it to millikelvin temperatures so that the superconducting properties are maintained. The processor is connected via microwave lines, control electronics, and readout systems that generate and detect the qubits’ quantum states. Additional components like cryogenic amplifiers, shielding, and classical control hardware are included to stabilize and interface with the quantum processor. 

\end{document}